%% file: JChemPhysReview.tex
\documentclass[aip,jcp,reprint]{revtex4-1}
\usepackage[T1]{fontenc}
\usepackage[latin9]{inputenc}
\usepackage{amsmath}
\usepackage{graphicx}
\usepackage{amssymb}
\usepackage{hyperref}
\include{JPSCommandsJCP}

\newcommand{\DKL}{D_{\mathrm{KL}}}

\begin{document}

\title{Sloppiness and Emergent Theories in Physics, Biology, and Beyond}

\author{Mark K.~Transtrum}
\affiliation{Department of Physics and Astronomy, Brigham Young University, Provo, Utah 84602, USA}

\author{Benjamin Machta}
\affiliation{Lewis-Sigler Institute for Integrative Genomics, Princeton University, Princeton, NJ, USA}

\author{Kevin Brown}
\affiliation{Departments of Biomedical Engineering, Physics, Chemical and Biomolecular Engineering, and Marine Sciences, University of Connecticut, Storrs, CT, USA}
\affiliation{Institute for Systems Genomics, University of Connecticut, Storrs, CT, USA}

\author{Bryan C.~Daniels}
\affiliation{Center for Complexity and Collective Computation, Wisconsin Institute for Discovery, University of Wisconsin, Madison, WI, USA}

\author{Christopher R. Myers}
\affiliation{Laboratory of Atomic and Solid State Physics, Cornell University, Ithaca, NY, USA}
\affiliation{Institute of Biotechnology, Cornell University, Ithaca, NY, USA}

\author{James Sethna}
\affiliation{Laboratory of Atomic and Solid State Physics, Cornell University, Ithaca, NY, USA}

\begin{abstract}
  Large scale models of physical phenomena demand the development of
  new statistical and computational tools in order to be effective.
  Many such models are `sloppy', i.e., exhibit behavior controlled by
  a relatively small number of parameter combinations.  We review an
  information theoretic framework for analyzing sloppy models.  This
  formalism is based on the Fisher Information Matrix, which we
  interpret as a Riemannian metric on a parameterized space of models.
  Distance in this space is a measure of how distinguishable two
  models are based on their predictions.  Sloppy model manifolds are
  bounded with a hierarchy of widths and extrinsic curvatures.  We
  show how the manifold boundary approximation can extract the simple,
  hidden theory from complicated sloppy models.  We attribute the
  success of simple effective models in physics as likewise emerging
  from complicated processes exhibiting a low effective
  dimensionality.  We discuss the ramifications and consequences of
  sloppy models for biochemistry and science more generally.  We
  suggest that the reason our complex world is understandable is due
  to the same fundamental reason: simple theories of macroscopic
  behavior are hidden inside complicated microscopic processes.
\end{abstract}

 
\maketitle

\section{Parameter indeterminacy and sloppiness}



As a young physicist, Freeman Dyson paid a visit to Enrico Fermi~\cite{Dyson:2004} 
(recounted in~\citet{Loew13}). Dyson wanted to tell Fermi
about a set of calculations that he was quite excited about.  Fermi asked Dyson
how many parameters needed to be tuned in the theory to match
experimental data.  When Dyson replied there were four,
Fermi shared with Dyson a favorite adage of his that
he had learned from Von Neumann: ``with four parameters I can fit an
elephant, and with five I can make him wiggle his trunk.''  Dejected, Dyson took
the next bus back to Ithaca.

As scientists, we are frequently in a similar position to Dyson.  We
are often confronted with a model --- a heavily parameterized,
possibly incomplete or inaccurate mathematical representation of
nature --- rather than a theory (e.~g., the Navier-Stokes equations)
with few to no free parameters to tune.  In recent decades, fueled by
advances in computing capabilities, the size and scope of mathematical
models has exploded.  Massive complex models describing everything
from biochemical reaction networks to climate to economics are now a
centerpiece of scientific inquiry.  The complexity of these models
raises a number of challenges and questions, both technical and
profound, and demands development of new statistical and
computational tools to effectively use such models. 


Here we review several developments that have occurred in
the domain of sloppy model research.  {\em Sloppy} is the term used to
describe a class of complex models exhibiting large parameter
uncertainty when fit to data.  Sloppy models were initially
characterized in complex biochemical reaction
networks~\cite{BrownS03,BrownHCMLSC04}, but were soon afterward found
in a much larger class of phenomena including quantum Monte
Carlo~\cite{WaterfallCGBMBES06}, empirical atomic
potentials~\cite{FrederiksenJBS04}, particle accelerator
design~\cite{GutenkunstPhD}, insect flight~\cite{BermanW07},
and critical phenomena~\cite{MachtaCTS13}.

As a prototypical example, consider fitting decay data to a sum of
exponentials with unknown decay rates:
\begin{equation}
y(t, \theta) = \sum_\mu e^{-\theta_\mu t}.
\end{equation}
We denote the vector of unknown parameters by $\theta$.  These
parameters are to be inferred from data, for example, by nonlinear least squares.  
This inference problem is notoriously
difficult~\cite{ruhe1980fitting}.  Intuitively, we can understand why
by noting that the effect of each \emph{individual} parameter is
obscured by our choice to observe only the sum.  Parameters have
compensatory effects relative to the system's \emph{collective}
behavior.  A single decay rate can be decreased, for example, provided
other rates are appropriately increased to compensate.

This uncertainty can be quantified using statistical methods, as we
detail in section~\ref{sec:MathFramework}.  In particular, the Fisher
Information Matrix (FIM) can be used to estimate the uncertainty in
each parameter in our model.  The result for the sum of exponentials
is that each parameter is \emph{almost completely undetermined}.  Any
parameter can be varied by an infinite amount and the model could
still fit the data.
This does not mean that all parameters can be varied independently of
the others.  Indeed, while the statistical uncertainty in each
individual parameter might be infinite, the data places constraints on
\textit{combinations} of the parameters.

\begin{figure}
\includegraphics[width=\linewidth]{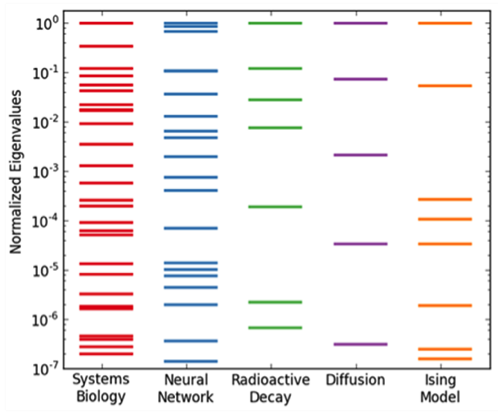}
\caption{\label{fig:Eigenvalues} \textbf{Sloppy eigenvalue spectra} 
  of multiparameter models from various
  fields~\cite{BrownS03,BrownHCMLSC04,WaterfallCGBMBES06,TranstrumMS11,MachtaCTS13}.
  Eigenvalues of the FIM, indicating sensitivity to perturbations along
  orthogonal directions in parameter space, are roughly evenly spaced
  in log-space, extending over many orders of magnitude.}
\end{figure}

The eigenvalues of the FIM tell us which parameter combinations
are well-constrained by the data and which are not.  Most of the FIM
eigenvalues are very small, corresponding to combinations of
parameters that have little effect on model behavior.  These
unimportant parameter combinations are designated \emph{sloppy}.  A
small number of eigenvalues are relatively large, revealing the few
parameter combinations that are important to the model (known as
\emph{stiff}).  It is generally observed that the FIM eigenvalues
decay roughly log-linearly, with each parameter combination being less
important than the previous by a fixed factor, as in
Figure~\ref{fig:Eigenvalues}.  Consequently there is not a
well-defined boundary between the stiff and sloppy combinations, and
four parameters really can ``fit the elephant''.

The degree of parameter indeterminacy in the simple
sum-of-exponentials model has been seen in many complex models of real
life systems for many of the same reasons.  The FIMs for seventeen
systems biology models have been shown to have the same characteristic
eigenvalue structure~\cite{GutenkunstWCBMS07}, and examples from other
scientific domains abound~\cite{WaterfallCGBMBES06}.  In each case,
observations measure a system's collective behavior,
and this means that when parameters have compensatory effects they
cannot be individually identified.

The ubiquity of sloppiness would seem to limit the usefulness of
complex parameterized models.  If we cannot accurately know parameter values, how
can a model be predictive?  Surprisingly, predictions are possible without precise 
parameter knowledge. As long as the model predictions depend on the same stiff
parameter combinations as the data, the predictions of the model will
be constrained in spite of large numbers of poorly determined parameters.

The existence of a few stiff parameter combinations can be understood
as a type of {\em low effective dimensionality} of the model.  In
section~\ref{sec:WhySloppy} we make this idea quantitative by
considering a geometric interpretation of statistics.  This leads
naturally to a new method of model reduction that constructs
low-dimensional approximations to high-dimensional models
(section~\ref{sec:ModelReduction}).  These low-dimensional
approximations are useful for revealing the emergent control
mechanisms that govern the system's behavior, i.e., extracting a
simple emergent theory of the collective behavior from the larger, complex
model.

Simple approximations to complex processes are common in physics
(section~\ref{sec:Physics}).  The ubiquity of sloppiness suggests that
similarly simple models can be constructed for other complex systems.
Indeed, sloppiness has a number of profound implications for the
unreasonable effectiveness of mathematics~\cite{Wigner60} and the
hierarchical structure of scientific theories~\cite{anderson1972more}.
We discuss some of these consequences specifically for modeling
biochemical networks in section~\ref{sec:biochemical}.  We discuss
more generally the implications of sloppiness for mathematical
modeling in section~\ref{sec:Consequences}.  We argue that sloppiness
is the underlying reason why the universe (a complete description of
which would be indescribably complex) is comprehensible.


\section{Mathematical Framework}
\label{sec:MathFramework}
In this section we use information
theory
to define key measures of sloppiness geometrically~\cite{Amari00}.  We
first consider the special case of model selection for models fit to
data by least squares.  We then generalize to the case of arbitrary
probabilistic models.  The key insight is that the Fisher Information
defines a Riemannian geometry on the space of possible
models~\cite{Amari00}.
The geometric picture allows us to show (in section
\ref{sec:WhySloppy}) that this local sloppy structure in the metric is
paralleled by a global hyper-ribbon structure of the entire space of
possible models.

\begin{figure}
\includegraphics[width=\linewidth]{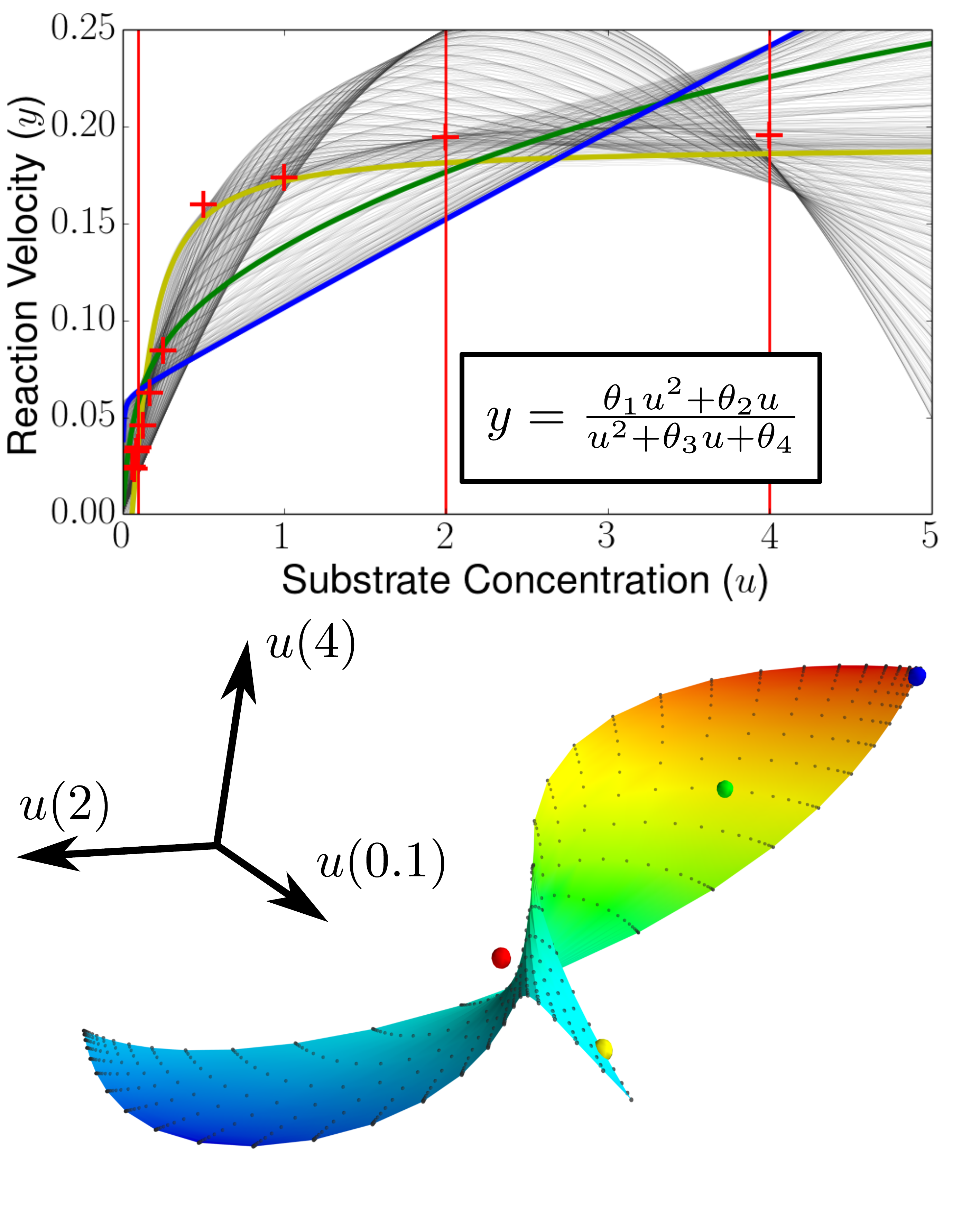}
\caption{\label{fig:AERMM} \textbf{The Model Manifold} 
  A simple model~\cite{Averick1992,kowalik1968analysis}
  of an enzyme-catalyzed reaction can be expressed as a rational
  function in substrate concentration ($u$) with four parameters
  ($\theta$) predicting the reaction velocity ($y$) (inset, top). By varying
  $\theta$ the model can predict a variety of behaviors $y$ as a function
  of $u$ (top). The
  model manifold is constructed by collecting all possible predictions
  of the model at specific values of $u$ (red vertical lines at $u =
  0.1, 2.0, 4.0$). To visualize the manifold, we take a two-dimensional
  cross section of the four dimensional manifold by choosing
  $\theta_1$ and $\theta_2$ to best fit the experimental data.  Varying
  $\theta_3$ and $\theta_4$ then maps out a two-dimensional surface
  of possible values in three-dimensional data space (bottom).  Each
  curve in the top figure corresponds to a point of the same color
  on the model manifold (bottom); the red crosses on top are data corresponding
  to the red dot below.}
\end{figure}

We begin with a simple case -- a model $\by$ predicting data $\bd$ at
experimental conditions $\bu$, with independent Gaussian errors; each
of these are vectors whose length $M$ is given by the number of data
points.  Our model depends on $N$ parameters $\btheta$.  In general,
an arbitrary model is a mathematical mapping from a parameter space
into predictions, so interpreting a model as a manifold of dimension
$N$ embedded in a data space $\Rbb^M$ is natural; the parameters
$\btheta$ then become the coordinates for the manifold. If our error
bars are independent and Gaussian all with the same width (say,
$\sigma=1$), finding the best fit of model to data is a least squares
data fitting problem, as we illustrate in Figure~\ref{fig:AERMM}.  In
this case, we assume that each experimental data point, $d_i$, is
generated from a parameterized model, $y(u_i, \theta)$, plus random
Gaussian noise, $\xi_i$:
\begin{equation}
  \label{eq:modelnoisedef}
  d_i = y(u_i, \theta) + \xi_i.
\end{equation}
Since the noise is Gaussian,
\begin{equation}
  \label{eq:gaussianpdf}
  P(\xi) \propto e^{-\xi^2/2},
\end{equation}
maximizing the log likelihood is equivalent to minimizing the sum of squared residuals, sometimes referred to as the cost or $\chi^2$ function:
\begin{equation}
  \label{eq:cost}
  \chi^2(\theta) = \sum_i r_i^2 = \sum_i \left( d_i - y(u_i, \theta) \right)^2.
\end{equation}

A sum of squares is reminiscent of a Euclidean distance.  Fitting a
model to data by least squares is therefore minimizing a distance in
data space between the observed data and the model.  Distance in data
space measures the quality of a fit to experimental data (red point in
Figure~\ref{fig:AERMM}).  Distance on the manifold is induced by,
i.e., is the same as, the corresponding distance in data space and
is measured using the metric tensor
\begin{equation}
  \label{eq:FIMNLLS}
  g_{\mu\nu} = \sum_i \frac{\partial y(u_i, \theta)}{\partial \theta^\mu} \frac{\partial y(u_i, \theta)}{\partial \theta^\nu} = (J^TJ)_{\mu\nu},
\end{equation}
where $J_{i\mu} = \partial y(u_i, \theta) / \partial \theta^\mu$ is
the Jacobian matrix of partial derivatives. This metric tensor is precisely
the Fisher Information Matrix (FIM) defined below, specialized to our 
least-squares problem. It is the least squares
Hessian matrix of second derivatives of $\half \chi^2$ from eqn~\ref{eq:cost},
evaluated where the data point $\bd$ is taken to be perfectly predicted
by $\by(\btheta)$.
On the manifold, distance is a
measure of identifiability -- how difficult it would be to distinguish
nearby points on the manifold (i.e., alternate models) through their
predictions.

We can generalize from this least-squares fitting problem to encompass
other models (like the Ising model) where the predictions are for entire
probability distributions.
For the purpose of modeling, the output of our
model is a probability distribution for $x$, the outcome of an
experiment.  A parameterized space of models is thus defined by
$P(x | \theta)$.  To define a geometry on this space we must define a
measure of how distinct two points $\theta_1$ and $\theta_2$ in
parameter space are, based on their {\em predictions}.  Imagine
getting a sequence of assumed independent data $x_1,x_2,...$ with the
task of inferring the model which produced them.  The likelihood that
model $\theta_1$ would have produced this data is given by
\begin{equation}
P(x_1,x_2,...|\theta)=\prod_i P(x_i|\theta) = \exp{\left(\sum_i \log{P(x_i|\theta)}\right)}.
\end{equation}

In maximum likelihood estimation our goal is simply to find the
parameter set $\theta$ which maximizes this likelihood.  It is
useful to talk about $\log{P(x|\theta)}$, the log-likelihood, as
this is the unique measure which is additive for independent data
points.  The familiar Shannon entropy of a model's predictions
$x$ is given by minus the expectation value of the
log-likelihood:
\begin{equation}
S(\theta)=-\sum_{x}P(x|\theta)\log{P(x|\theta)}.
\end{equation}
We can also define an analogous quantity that measures
the likelihood that model $\theta_2$
would produce typical data from $\theta_1$:
\begin{equation}
\sum_{x}P(x|\theta_1)\log{P(x|\theta_2)}.
\end{equation}
The Kullback-Leibler divergence between $\theta_1$ and
$\theta_2$ measures how much more likely $\theta_1$ is to
produce typical data from $\theta_1$ than $\theta_2$ would
be:
\begin{equation}
\DKL(\theta_1||\theta_2)=\sum_{x}P(x|\theta_1)\big(\log{P(x|\theta_1)}-\log{P(x|\theta_2)}\big).
\end{equation}
Thus $\DKL$ is an intrinsic measure of how difficult distinguishing
these two models will be from their data.

The KL divergence does not satisfy the mathematical requirements of a
distance measure.  It is asymmetric, and does not satisfy even a weak
triangle inequality: In some cases
$\DKL(\theta_1||\theta_3)>\DKL(\theta_1||\theta_2)+\DKL(\theta_2||\theta_3)$.
However, for models whose parameters $\theta$ and
$\theta+\delta \theta$ are quite close to one another, the
leading terms are symmetric and can be written as:
\begin{equation}
\DKL(\theta||\theta+\delta \theta) = g_{\mu\nu} \delta \theta^\mu \delta \theta^\nu + \mathcal{O} \delta \theta^3
\end{equation} 
 where $g_{\mu\nu}$ is the Fisher Information Matrix (FIM), which can be written:
\begin{equation}
\label{eq:defFisher}
g_{\mu\nu}(P_\theta)=-\sum_{x} P_\theta(x) \frac{\partial}{\partial \theta^\mu}\frac{\partial}{\partial \theta^\nu}\log P_\theta(x).
\end{equation}
The FIM has all the properties of a metric tensor.  It is symmetric
and positive semi-definite (because no model can on average be better
described by a different model) and it transforms properly under a
coordinate reparameterization of $\theta$.  Information
Geometry~\cite{beale1960confidence,bates1980relative,amari1985differential,amari1987differential,murray1993differential,amari2007methods,TranstrumMS10,TranstrumMS11}
is the study of the properties of the model manifold defined by this
metric.  In particular, it defines a space of models in a way that
does not depend on the labels given to the parameters, which are
presumably arbitrary; should one measure rate constants in seconds or
hours, and more problematically, should one label these constants as
rates, or time constants?  Information Geometry makes clear that some
aspects of a parameterized model can be defined in ways that are
invariant to these arbitrary choices.

%
%

\section{Why Sloppiness?  Information Geometry}
\label{sec:WhySloppy}

Sloppy models can be identified by the characteristic eigenvalue
spectrum of the FIM.  We interpret the existence of many small
eigenvalues in the FIM to be representative of a complex model with a
low effective dimensionality.  Many combinations of parameters
have minimal effect on the behavior of the model, while the key
features of model behavior are controlled by a relatively small number
of stiff parameter combinations.  In a sense, then, there really is a
simpler `theory' embedded in the multiparameter `model'.

In this section we make the notion of low effective dimensionality
explicit.  We will see that although this interpretation of sloppy
models turns out to be correct, the eigenvalues of the FIM are not
sufficient to make this conclusion.  Instead, we use the geometric
interpretation of modeling introduced in
section~\ref{sec:MathFramework} that allows us to quantify important
features of the model in a global and parameterization independent
way.  The effort to develop this formalism will pay further dividends
when we consider model reduction in section~\ref{sec:ModelReduction}.

To understand the limitations of interpreting the eigenvalues of the
FIM, we return to the question of model reparameterization.  Something
as trivial as changing the units of a rate constant from Hz to kHz
changes the corresponding row and column of the FIM by a factor of
1000, in turn changing the eigenvalues.  Of course, none of the model
predictions are altered by such a change since a correcting factor of
1000 will be introduced throughout the model.  More generally, the FIM
can be transformed into any positive definite matrix by a simple
linear transformation of parameters while model predictions are always
invariant to such a reparameterization.

Although the FIM eigenvalues are not invariant to reparameterization,
we can use information geometry to search for a parameterization
independent measure of sloppiness.  With the definitions of
section~\ref{sec:MathFramework}, computational differential geometry
can be used to explore the a wide variety of model manifolds in
a parameter independent way.  A review of these methods is beyond the
scope of this paper, and we refer the interested reader to
references~\cite{TranstrumMS10,TranstrumMS11} or any standard text on
differential geometry~\cite{spivak1979,Ivancevic2007}.

The key geometric feature of the model manifolds of nonlinear sloppy
systems is that they have {\em boundaries}. Many parameters and
parameter combinations can be taken to extreme values (zero or
infinity) without leading to infinite predictions. These boundaries
can be explored by numerically constructing manifold geodesics:
analogs of straight lines on curved surfaces.  The arc lengths of
geodesics are a measure of the width of the model manifold in each
direction.  Measuring these arc lengths for a sloppy model shows that
the widths of sloppy model manifolds are exponentially distributed,
reminiscent of the exponential distribution of FIM eigenvalues.
Indeed, when we use dimensionless model parameters
(e.~g.~log-parameters), the square roots of the FIM eigenvalues are a
reliable approximation to the widths of the manifold in the
corresponding eigendirections~\cite{TranstrumMS10,TranstrumMS11}.

\begin{figure}
\includegraphics[width=\linewidth]{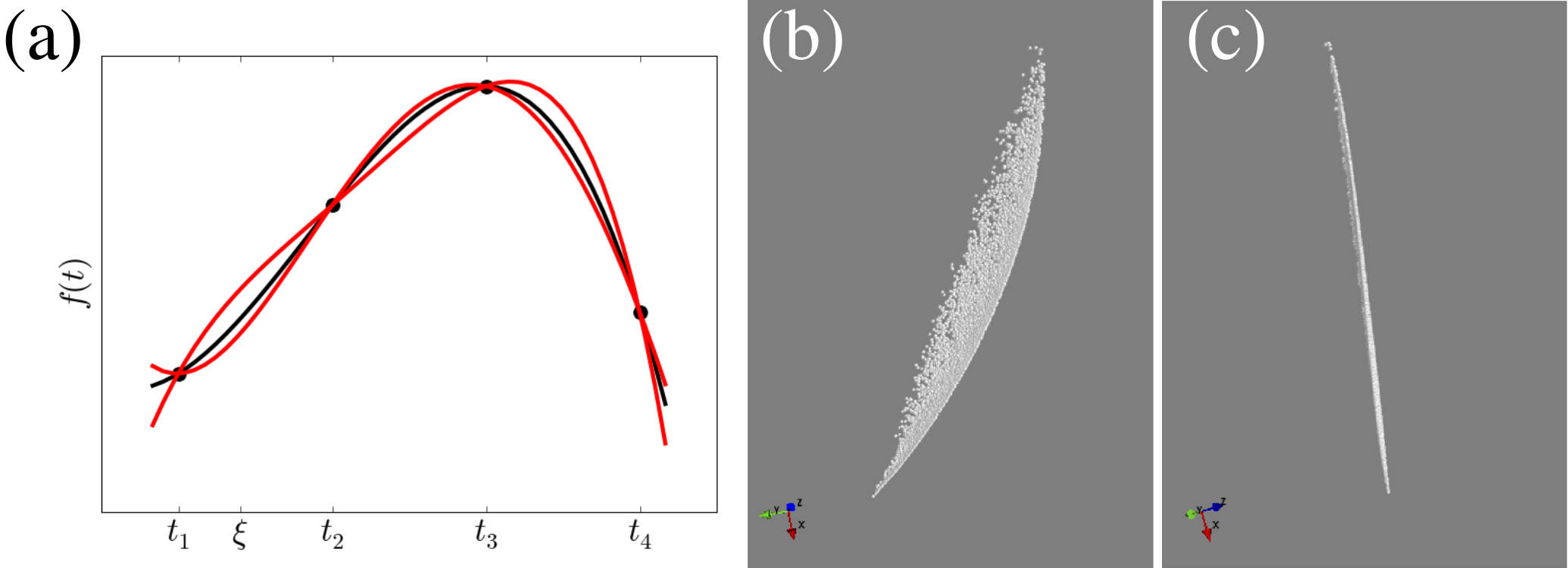}
\caption{\label{fig:Hyperribbon} \textbf{Hyperribbon.}
(a)~Given a multiparameter model for
one-dimensional data $f(t)$ at different times $t$, the model manifold
has a different dimension for every time $t$. Specifying a data point
$f(t_n)$ thus gives a cross-section of the model manifold, and also
reduces the uncertainty in the values of neighboring points -- hence
giving the cross-section a narrower width -- a hyperribbon. Interpolation
theory~\cite{TranstrumMS10,TranstrumMS11} can be used to quantify this
qualitative argument. 
(Figure from fig.~5 of~\cite{TranstrumMS11}.) 
(b,c)~Two views of a hyperribbon cross section of a model manifold. The
model is decaying exponentials fit to radioactive decay
data~\cite{WaterfallCGBMBES06}. Notice the ribbon-like structure of this
three-dimensional projection: long, narrow, and very thin.}
\end{figure}

The exponential distribution of manifold widths has been described as
a \emph{hyperribbon} (Fig.~\ref{fig:Hyperribbon}).
A three-dimensional ribbon has a long
dimension, a broad dimension, and a very thin dimension.  The observed
hierarchy of exponentially decreasing manifold widths are a
high-dimensional generalization of this structure.  We will explore
the nature of these boundaries in more detail when we discuss model
reduction in section~\ref{sec:ModelReduction}.

The observed hierarchy of widths can be demonstrated analytically for
the case of a single independent variable (such as time or substrate
concentration in Figure~\ref{fig:AERMM}) by appealing to theorems for
the convergence of interpolating functions (Fig.~\ref{fig:Hyperribbon}(a)).
Consider removing a few
degrees of freedom from a time series by fixing the output of the
model at a few time points.  The resulting model manifold corresponds
to a cross-section of the original.  Next, consider how much the
predictions at intermediate time points can be made to vary as the
remaining parameters are scanned.  As more and more predictions are
fixed (i.e., considering higher-dimensional cross sections of the
model manifold), we intuitively predict that the behavior of the model
at intermediate time points will become more constrained.
Interpolation theorems make this intuition formal; presuming
smoothness or analyticity of the predictions as a function of time,
one can demonstrate an exponential hierarchy of widths consistent with
the hyperribbon structure observed
empirically~\cite{TranstrumMS10,TranstrumMS11}.

The exponential hierarchy of manifold widths makes explicit the notion
of a low effective dimensionality in the model that was hinted at by
the eigenvalues of the FIM.  It also helps illustrate how models can
be predictive without parameters being tightly constrained.  Only
those parameter combinations that are required to fit the key features
of the data need to be estimated accurately.  The remaining parameter
combinations (controlling for example the high-frequency behavior in
our time series example) are unnecessary.  In short, the model
essentially functions as an interpolation scheme among observed data
points.  Models are predictive with unconstrained parameters when the
predictions interpolate among observed data.

Understanding models as generalized interpolation schemes makes
additional predictions about the generic structure of sloppy model
manifolds.  Not only is there an exponential distribution of widths,
there is also an exponential distribution of extrinsic curvatures.
Furthermore, these curvatures are relatively small in relation to the
widths, making the model manifold surprisingly flat.  Most of the
nonlinearity of the model's parameters take the form of `parameter
effects
curvature'~\cite{bates1980relative,Bates1981,Bates1983,Bates1988},
(equivalent to the connection coefficients~\cite{TranstrumMS11}).  The
small extrinsic curvature of many models was a mystery first noted in
the early 1980s~\cite{bates1980relative} that is explained by
interpolation arguments.

\section{Model Reduction}
\label{sec:ModelReduction}

In this section, we leverage the power of the information geometry
formalism to answer the question: how can a simple effective model be
constructed from a (more-or-less) complete but sloppy representation
of a physical system?  Our goal is to construct a physically
meaningful representation that reveals the simple `theory' that is
hidden in the model.

The model reduction problem has a long history, and it would be impossible
in this review to even approach a comprehensive survey of literature on the subject.
Several standard methods have emerged that have proven effective in
appropriate contexts.  Examples include clustering components into
modules~\cite{wei1969lumping,liao1988lumping, huang2005systematic},
mean field theory, various limiting approximations (e.g., continuum,
thermodynamic, or singular limits), and the renormalization
group~\cite{goldenfeld1992lectures,zinn2007phase}.  Considerable effort
has been devoted by the control and engineering communities to
approximate large-scale dynamical
systems~\cite{saksena1984singular,kokotovic1999singular,
  naidu2002singular, antoulas2005approximation,lee2010multi}, leading
to the method of balanced
truncation~\cite{moore1981principal,dullerud2000course,
  gugercin2004survey}, including several structure preserving
variations~\cite{zhou1995structurally,li2005structured,sandberg2009model}
and generalizations to nonlinear
cases~\cite{scherpen1993balancing,lall2002subspace, krener2008reduced}.
Methods for inferring minimal dynamical models in cases for which
the underlying structure is not known are also beginning to be
developed~\cite{DanielsN14,DanielsN15}.

Unfortunately, many automatic methods produce `black box'
approximations.  For most scenarios of practical importance, a reduced
representation alone has limited utility since attempts to
engineer or control the system typically operate on the microscopic level.  For
example, mutations operate on individual genes and drugs target
specific proteins.  A method that explicitly reveals how microscopic
components are `compressed' into a few effective degrees of freedom
would be very useful.  On the other hand, methods that do explicitly
connect microscopic components to macroscopic behaviors have limited
scope since they often exploit special properties of the model's
functional form, such as symmetries.  Consider, for example, the
renormalization group, which operates on field theories with a scale
invariance or conformal symmetry.  Simplifying modular network
systems, such as biochemical networks, is particularly challenging
due to inhomogeneity and lack of symmetries.

The Manifold Boundary Approximation Method
(MBAM)~\cite{transtrum2014model} is an approach to model approximation
whose goal is to alleviate these challenges.  As the name implies, the
basic idea is to approximate a high-dimensional, but thin model
manifold by its boundary.  The procedure can be summarized as a four
step algorithm.  First, the least sensitive parameter combination is
identified from an eigenvalue decomposition of the FIM.  Second, a
geodesic on the model manifold is constructed numerically to identify
the boundary.  Third, having found the edge of the model manifold, the
corresponding model is identified as an approximation to the original
model.  Fourth, the parameter values for this approximate model are
calibrated by fitting the approximate model to the behavior of the
original model.

The result of this procedure is an approximate model that has one less
parameter and that is marginally less sloppy than the original.
Iterating the MBAM algorithm therefore produces a series of models of
decreasing complexity that explicitly connect the microscopic
components to the macroscopic behavior.  These models correspond to
hyper-corners of the original model manifold.  The method requires
only that the model manifold have a hierarchy of boundaries while
making no assumptions about the mathematical form of the model or
underlying physics of the system.  As such, MBAM is a very general
approximation scheme.

The key component that enables MBAM are the edges of the model
manifold.  The existence of these edges was first noted in the context
of data fitting~\cite{TranstrumMS10} and MCMC sampling of Bayesian
posterior distributions~\cite{GutenkunstPhD}.  It was noted that
algorithms would often `evaporate' parameters, i.e., allow them to
drift to extreme, usually infinite, values.  These extreme parameter
values correspond to limiting behaviors in the model, i.e., manifold
boundaries.

`Evaporated parameters' are especially problematic for numerical
algorithms.  Numerical methods often push parameters to the edge of
the manifold and then become lost in parameter space.  Consider
the case of MCMC sampling of a Bayesian posterior.  If a parameter
drifts to infinity, there is an infinite amount of entropy associated
with that region of parameter space and the sampling will never
converge.  Furthermore, the model behavior of such a region will
always dominate the posterior distribution~\cite{GutenkunstPhD}.

For data fitting algorithms, methods such as Levenberg-Marquardt
operate by fitting the key features of the data first (i.e., the
stiffest directions), followed by successive refining approximations
(i.e., progressively more sloppy components).  While fitting the
initial key features, algorithms often evaporate those parameters
associated with less prominent features of the data.  The algorithm is
then unable bring the parameters away from infinity in order to
further refine the fit~\cite{TranstrumMS10}.

\begin{figure}[t]
 \includegraphics[width=\columnwidth]{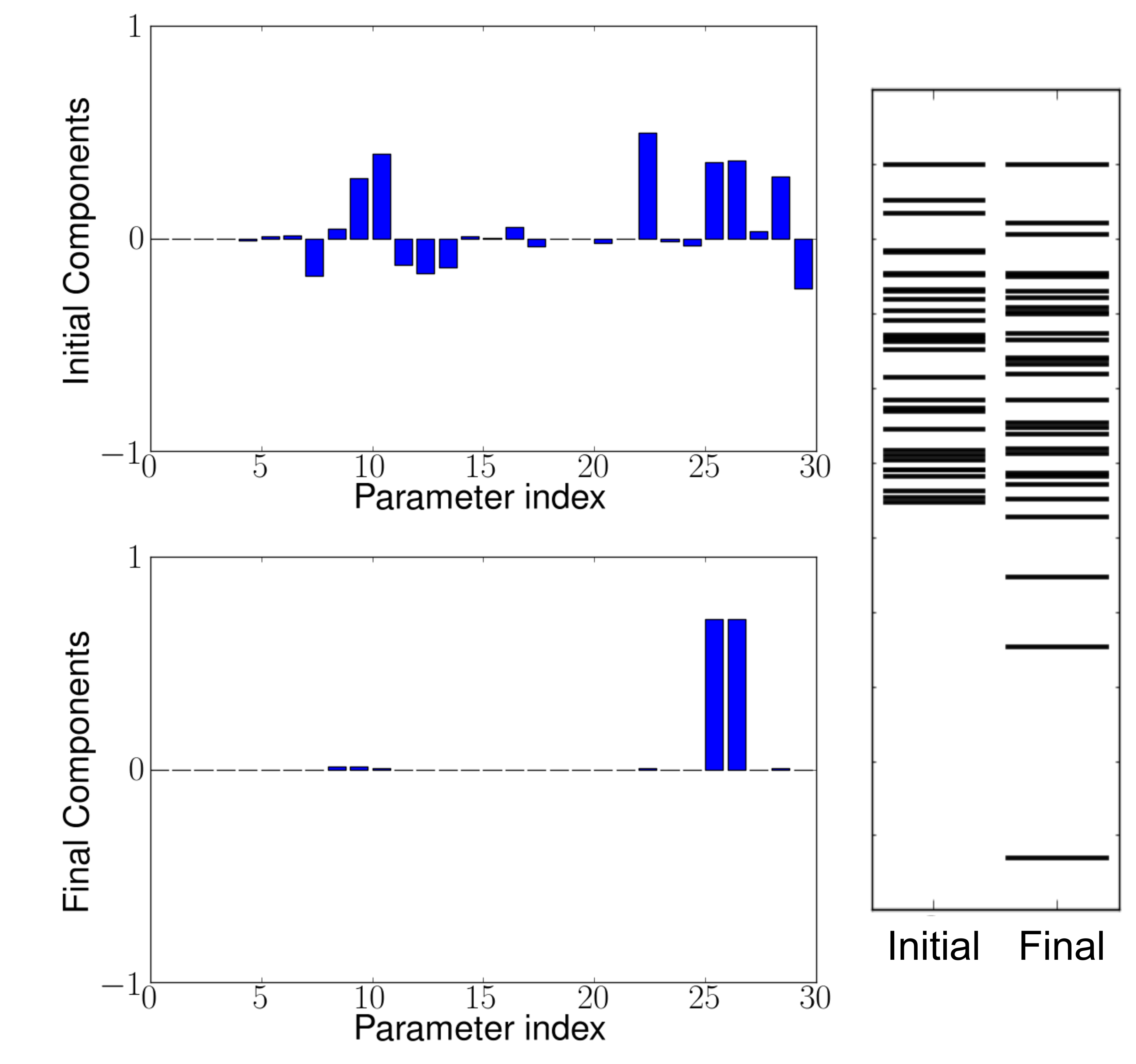}
 \caption{\label{fig:Components} \textbf{Identifying the boundary
     limit~\cite{transtrum2014model}}.
   The components of the smallest eigenvector of the FIM is
   often a complicated combination of bare parameters that is
   difficult to either interpret or remove from the model (top left).
   By following a geodesic to the manifold boundary, the combination
   rotates to reveal a limiting behavior (bottom left); here two
   parameters (a reaction rate and a Michaelis-Menten constant) become
   infinite. The limiting behavior is revealed when the smallest
   eigenvalue has become separated from the other eigenvalues (right).}
 \end{figure}

Although problematic for numerical algorithms, manifold edges are
useful for both approximating (ala MBAM) and interpreting complex
models.  To illustrate, we consider an EGFR signaling
model\cite{BrownS03}.  Figure~\ref{fig:Components} illustrates
components of one eigenparameter, corresponding in this case to the
smallest eigenvalue of the FIM.  Notice that the eigenparameters do
not align with bare parameters of the model, but typically involve an
unintuitive combination of bare parameters.  However, by following a
geodesic along the model manifold to the manifold edge (step 2 of the
MBAM algorithm), these complex combinations slowly rotate to reveal
relatively simple, interpretable combinations that correspond to a
limiting approximation of the model.  For example, the EGFR model in
reference~\cite{BrownS03} consists of a network of Michaelis-Menten
reactions.  The boundary revealed~\cite{transtrum2014model} in
Figure~\ref{fig:Components} corresponds to the limit of a reaction
rate and a Michaelis-Menten constant becoming infinite while their
ratio is finite:
\begin{eqnarray}
  \label{eq:Limit}
  \frac{d}{dt} [ A ] & = & \frac{k [A] [B]}{K_M + [A]} + \dots \\
 & \rightarrow & \left( \frac{k}{K_M} \right) [A] [B] + \dots,
\end{eqnarray}
where $[A]$ and $[B]$ are concentrations of two enzymes in the model
and the ratio $k/K_M$ is the renormalized parameter in the approximate
model.

\begin{figure}[t]
 \includegraphics[width=\columnwidth]{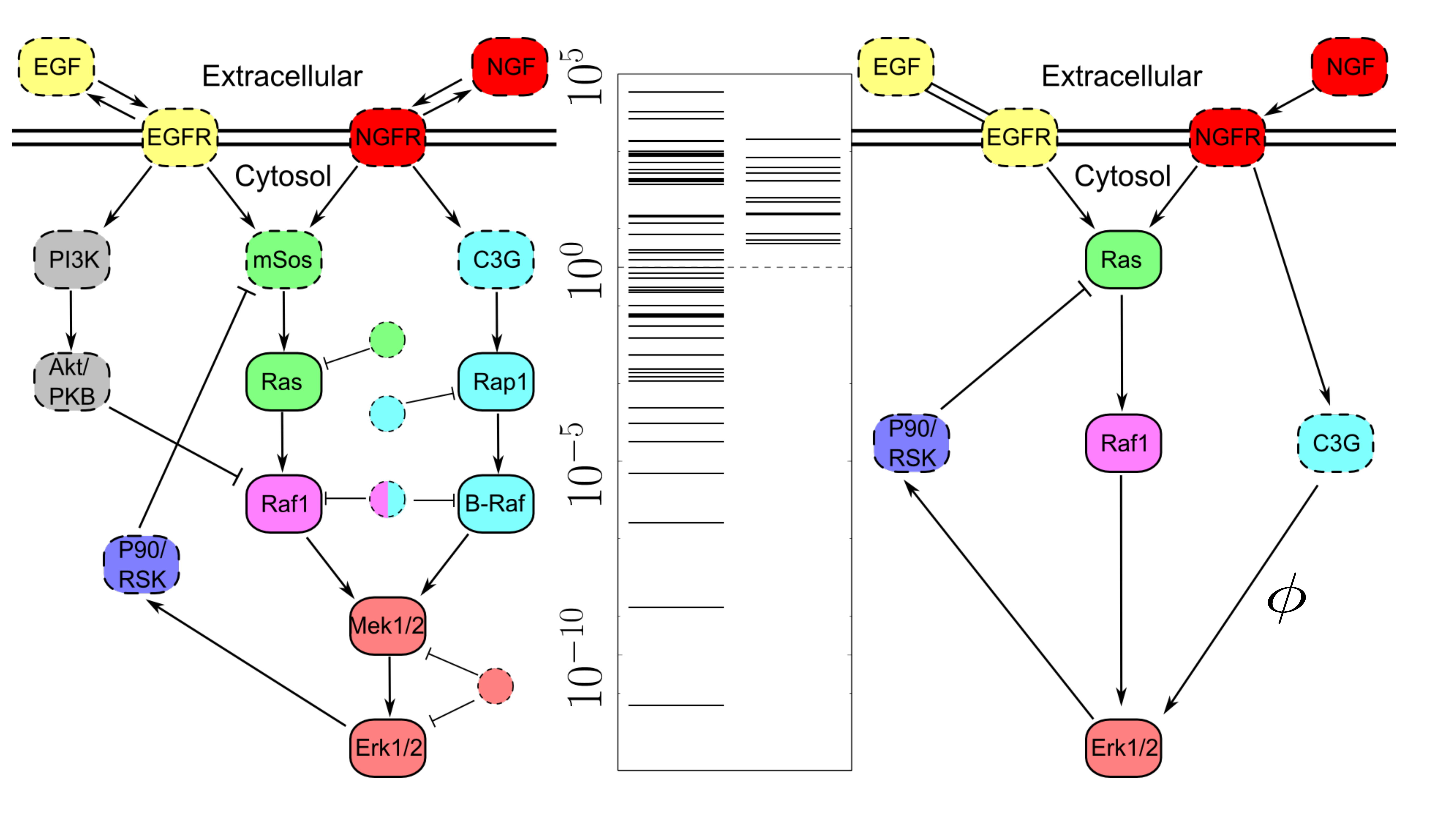}
 \caption{\label{fig:PC12reduced}\textbf{Original~\cite{BrownS03}
     and reduced~\cite{transtrum2014model} EGFR
     models}. The interactions of the EGFR signaling
   pathway~\cite{BrownS03,BrownHCMLSC04} are summarized in the leftmost
   network.  Solid circles are chemical species for which the
   experimental data was available to fit.  Manifold boundaries reduce
   the model to a form (right) capable of fitting the same data and
   making the same predictions as in the original
   references~\cite{BrownS03,BrownHCMLSC04}.  The FIM eigenvalues
   (center) indicate that the simplified model has removed the
   irrelevant parameters identified as eigenvalues less than $1$
   (dotted line) while retaining the original model's predictive
   power.}
 \end{figure}

 Because the manifold edges correspond to models that are simple
 approximations of the original, the MBAM can be used to iteratively
 construct simple representations of otherwise complex processes.  By
 combining several limiting approximations, simple insights into the
 system behavior emerge that were obfuscated by the original model's
 complexity.  Figure~\ref{fig:PC12reduced} compares network diagrams
 for the original and approximate EGFR models.  The original
 consists of 29 differential equations and 48 parameters, while the
 approximate consists of 6 differential equations and 12 parameters
 and is notably \emph{not sloppy}.

 Because the MBAM process explicitly connects models through a series
 of limiting approximations, the parameters of the reduced model can
 be identified with (nonlinear) combinations of parameters in the
 original model.  For example, one of the twelve variables in the reduced
 model of Fig.~\ref{fig:PC12reduced} is written as an explicit combination
 of seven `bare' parameters of the original model:
\begin{equation}
\label{eq:phi}
\phi = \frac{ (\mathrm{k_{Rap1ToBRaf}}) (\mathrm{K_{mdBRaf}}) (\mathrm{k_{pBRaf}}) (\mathrm{K_{mdMek}}) }{ (\mathrm{k_{dBRaf}}) (\mathrm{K_{mRap1ToBRaf}}) (\mathrm{k_{dMek}}) }.
\end{equation}
Expressions such as this explicitly identify which combinations of
microscopic parameters act as emergent control knobs for the system.

MBAM naturally includes many other approximation methods as special
cases~\cite{transtrum2014model}.  By an appropriate choice of
parameterization, it is both a natural language for model reduction and a
systematic method that in practice can be mostly automated.

The MBAM is a semi-global approximation method.  Manifold boundaries
are a non-local feature of the model.  However, MBAM only explores the
region of the manifold in the vicinity of a single hyper-corner.  More
generally, it is possible to identify \emph{all} of the edges of a
particular model (and by extension, all possible simplified models).
This analysis is known information
topology~\cite{transtrum2014information}.

\section{Emergence in Physics as sloppiness}
\label{sec:Physics}
Unlike in systems biology, physics is dominated by effective models and
theories whose forms are often deduced long before a microscopic theory is
available.
This is in large part due to the great success of continuum limit arguments
and Renormalization Group (RG) procedures in justifying the expectation
and deriving the form of simple emergent theories. These methods
show that many different multi-parameter microscopic theories typically 
collapse onto one coarse-grained model, with the complex microscopics
being summarized into just a
few `relevant' coarse-grained parameters. This explains why an effective
theory, or an oversimplified `cartoon' microscopic theory, can often make
quantitatively correct predictions.  Thus, while three dimensional
liquids have enormous microscopic diversity, in a certain regime (lengths
and times large compared to molecules and their vibration periods),
their behavior is determined 
entirely by their viscosity and density.  Although two different
liquids can be microscopically completely different, their effective
behavior is determined only by the projection of their microscopic details
onto these two control parameters.  This parameter space compression
underlies the success of renormalizable and continuum limit models.

This connection has been made explicit, by examining the FIM for typical
microscopic models in physics~\cite{MachtaCTS13}. A microscopic hopping
model for the continuum diffusion equation quickly develops `stiff' directions
corresponding to the parameters of the continuum theory -- the total number of
particles, net mean velocity, and diffusion constant. As
time evolves, all other microscopic parameter combinations become increasingly
sloppy -- irrelevant for prediction of long-time behavior.
Similarly, a microscopic long-range Ising model for ferromagnetism, when
observed on long length scales, develops stiff directions along precisely those
parameter combinations deemed `relevant' under the renormalization group. 


\begin{figure*}[t]
 \includegraphics[width=2\columnwidth]{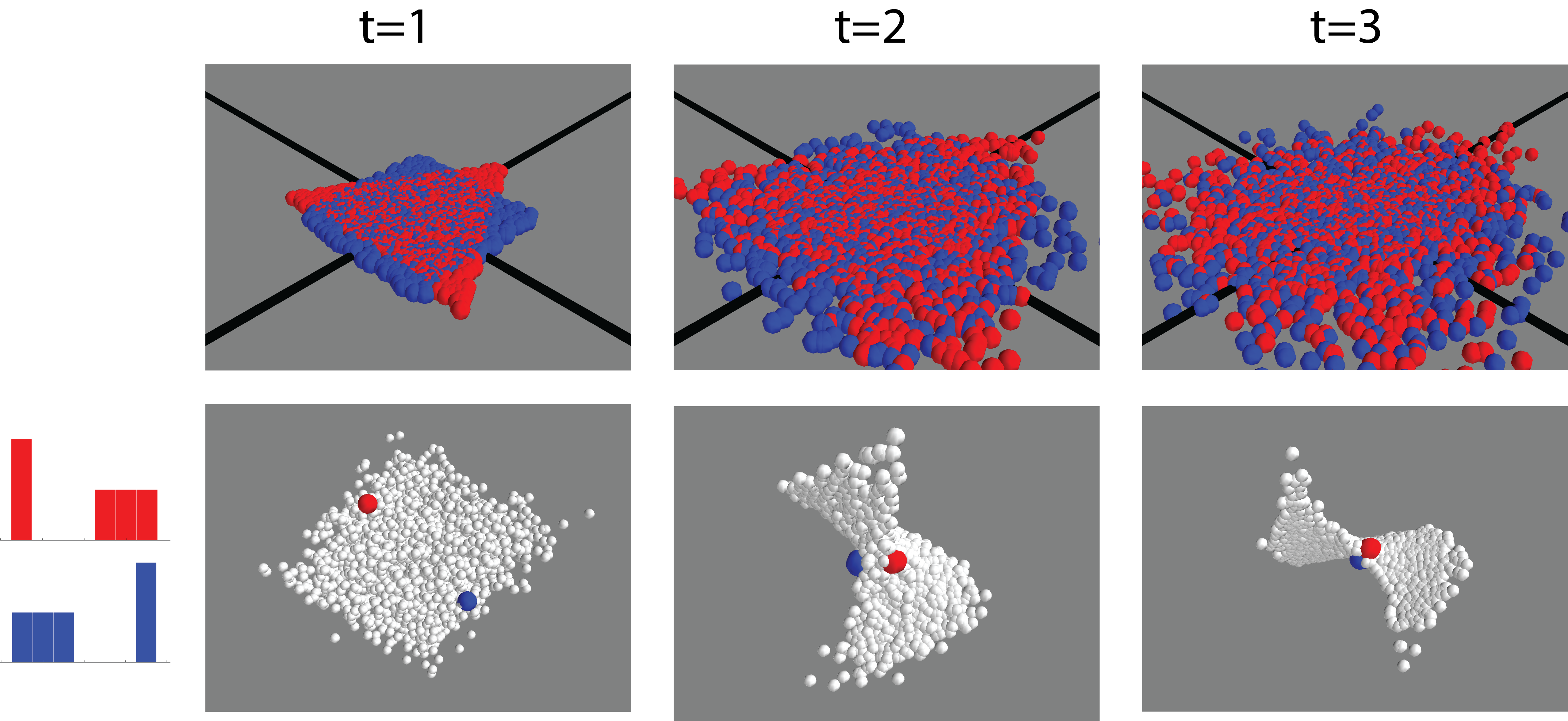}
 \caption{\label{fig:DiffRB}\textbf{Microscopic Motion becomes Diffusive. Top:} Simulated particles undergo stochastic motion in discrete time.  Red particles hop according to a triangular kernel, while blue particles hop according to a square kernel.  After a single time step, the particles have very different distributions in space, and neither resemble the distribution predicted by the diffusion equation.  However, as time evolves, most of the information about this kernel is lost.  Only the particles' diffusion tensor and average drift enter into a continuum description.  For the particles shown, the drift is 0, and their respective diffusion tensors are matched, so that the resulting distributions become quantifiably indistinguishable as time proceeds.  The compression of microscopic details mirrors the compression of molecular level detail in the emergence of diffusion as a continuum limit of motion in real fluids.
 \textbf{Bottom:} We consider the model manifold of a one dimensional lattice version of this diffusion example.  As with the triangle and square, the red and blue kernels shown on the left have drift 0, and identical second moments, though higher moments and the distribution in general are very different.  The remaining white points making up the manifold are taken from a uniform distribution in parameter space describing the probability of hopping to one of six nearest neighbors in a given time step, as in Ref.~\cite{MachtaCTS13}.  Here we plot a three dimensional projection of the model manifold taken from measurements of particle distributions at different time points.  After a single time step, this three dimensional projection from data space shows a `hyper-blob', with changes in parameters leading to a large diversity of observable behaviors.  In particular, the red and blue points are not close to each other even though their drifts are both 0 and their diffusion constants are matched;  as with the square and triangle, their distributions are easily distinguishable after a single time step.  However, after several stochastic steps, the model manifold takes on a hyperribbon structure.  Models for which all effective parameters are matched, like the red and blue kernels highlighted, rapidly move close to each other.  At late times, any model sufficiently flexible to capture the two remaining extended directions is adequate to describe effective behavior, explaining the ubiquity and success of the continuum diffusion equation.}
 \end{figure*}

Consider a model of stochastic motion as a stand-in for a molecular level description of particles moving through a possibly complicated fluid.  Such a fluid's properties depend on many parameters such as the bond angle of the molecules which make it up, all of which enter into the probability distribution for motion within the fluid, which can presumably be microscopically complex.  However, the law of large numbers says that as many of these random steps are added together, the long-time movement of particles will lead to them being distributed in space according to a Gaussian.  As this happens, diverse microscopic details must become compressed into the two parameters of a Gaussian distribution- its mean and width. As a concrete example, in the top of Figure~\ref{fig:DiffRB}, two very different microscopic motions are considered.  In each time step, red particles take a random step from a triangular distribution, while blue particles step according to a square distribution.  While these motions lead to very different distributions after a single time step, as time proceeds they become indistinguishable precisely because their first and second moments are matched.

This indistinguishability can be quantified by considering the model manifold of possible microscopic models of stochastic motion, again paralleling real fluids that can be microscopically diverse.  When probed at the intrinsic time and length scales of these fluids, we should make few assumptions about the type of motion we expect; in particular, we should allow for behaviors much more complicated than diffusion, by analogy with square and triangle described in two dimensions above.  Following Ref.~\cite{MachtaCTS13}, we consider a one dimensional `molecular level' model for stochastic motion, in which parameters describe the rates at which a particle hops to one of its close-by neighbors.  After a single time step, the corresponding model manifold is a `hyper-blob' (fig.~\ref{fig:DiffRB}, bottom) and two particular models, marked in red and blue, are distinguishable; they are not nearby on the model manifold.  The prediction space of a model is truly multidimensional in this regime- it cannot be described by the two parameter diffusion equation.  In this `ballistic' regime, motion is not described by the diffusion equation, and is presumably not just different, but genuinely more complicated.  However, as time proceeds, the model manifold contracts onto a hyper-ribbon, in which just two parameter combinations distinguish behavior.  In this regime, all points lie close to the two dimensional manifold predicted by the diffusion equation, and the red and blue points have become indistinguishable; they are now in close proximity on the manifold.

Using information geometry, approximations analogous to continuum
limits or the renormalization group can be found and used to construct
similarly simple theories in fields for which effective theories have
historically been difficult to construct or justify.

\section{Ramifications of Sloppiness in Biochemical Modeling}
\label{sec:biochemical}

\begin{figure*}[t]
 \includegraphics[width=2\columnwidth]{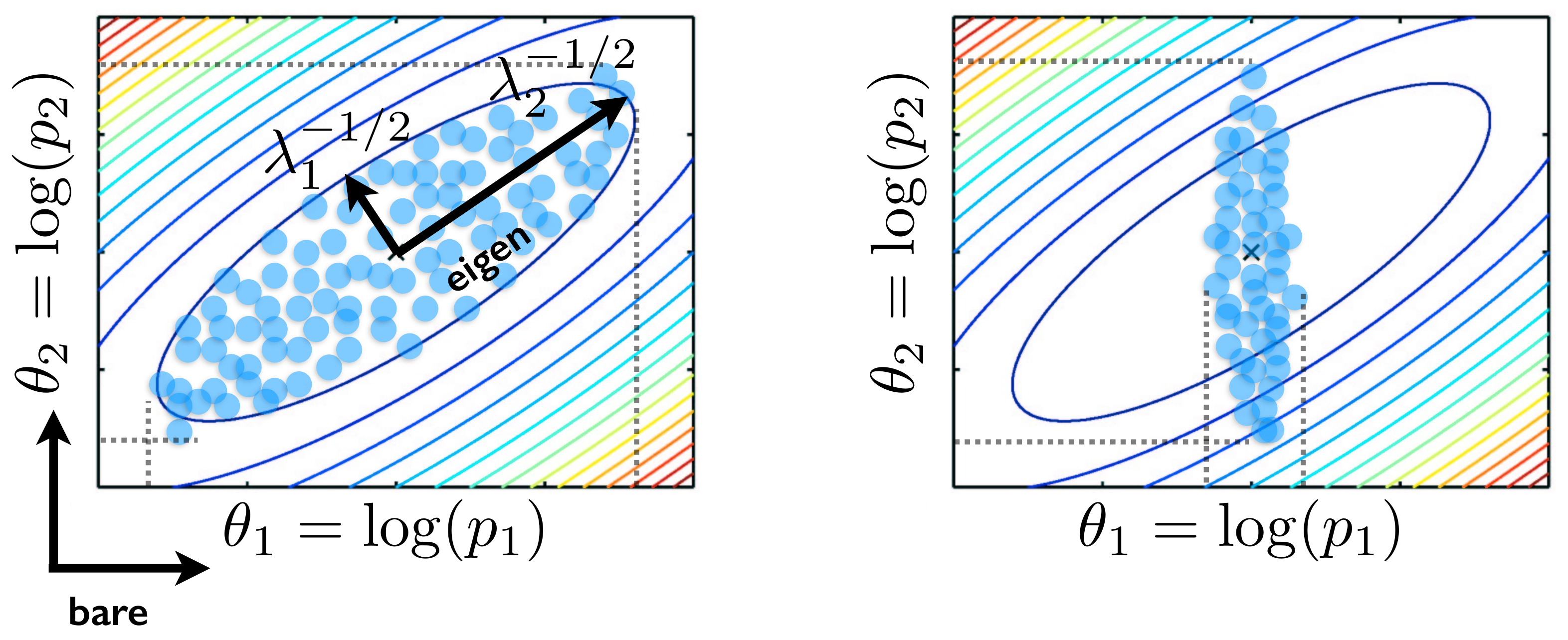}
 \caption{\label{fig:Ellipses}\textbf{Sloppiness in Parameter Space}. \textbf{Left:} A schematic of a typical Sloppy Model ensemble, pictured in two dimensions for clarity.  The underlying cost surface (with constant cost contours illustrated as ellipses) is generated by the fit to data.  Eigenvectors of the FIM correspond to principal axes of the ellipse, with widths of the ellipse inversely proportional to the square roots of the corresponding eigenvalues $\lambda_i$.  Points inside the ellipse each represent a set of parameters that fits the data within a given tolerance (in practice often created using a Monte Carlo approach), forming an ensemble representing uncertainty about the true values of parameters.  Sloppiness can result in good fits to data despite enormous uncertainties in `bare' $\theta$ parameters (dotted lines intersecting the axes). \textbf{Right:} Careful measurements of individual parameters (like $\theta_1$) can shrink uncertainty, but if even a single parameter remains unknown (like $\theta_2$), large predictive uncertainty can still result.}
 \end{figure*}

In previous sections, we have emphasized picturing the model manifold
in data space, as in Figure~\ref{fig:AERMM}; here, thin, sloppy dimensions
of the hyperribbon correspond to behavior that is minimally
dependent on parameters.  The dual picture in parameter space,
sketched in Figure~\ref{fig:Ellipses}, is one in which the set of parameters
that sufficiently well fit some given data is stretched to extend far
along sloppy dimensions.  This picture is important to understanding
implications for biochemical modeling with regard to parameter
uncertainty.


For instance, using the full EGFR signal transduction network
(left side of Figure~\ref{fig:PC12reduced}),
we may wish to make a prediction about an unmeasured experimental
condition, e.g.~the time-course of ERK activity upon EGF stimulation.
In general, if there are large uncertainties about the model's parameters,
we expect our uncertainty about this time-course to also be large.

Indeed this is the case if we neglect effects of compensation
among parameters and assume that uncertainties in
different parameters are uncorrelated.  If we view the problem of
uncertainties in model predictions
as coming from a lack of precision measurements of individual
parameters, we may try to carefully and independently measure
each parameter.  This can work if such measurements are
feasible, but can fail if even one relevant parameter remains unknown:
as in the right plot of Figure~\ref{fig:Ellipses},
a large uncertainty along the direction of a single parameter
corresponds to large changes in system-level behavior,
leading to large predictive uncertainties~\cite{GutenkunstWCBMS07}.

Contrasting with this approach, we can instead constrain the
model parameters with system-level measurements that are similar
to the types of measurements we wish to predict.  Due to the
phenomenon of sloppiness, we expect that this approach will
produce a subspace of acceptable parameters that will
include large uncertainties in individual parameter values
(left plot of Figure~\ref{fig:Ellipses}).
Again, this arises because two parameters that change the output in a
correlated way can consequently be simultaneously varied
without changing the model output.
It is often true that variance of system-level measurements
over the large sloppy parameter subspace is as small as
would require extremely precise measurements if parameters
were measured independently.
Predictions of interest can be made without precisely
knowing any single parameter.

Thus, in the sense that estimating parameters entails
discovering their precise values,
in sloppy models parameter estimation becomes useless.
This does not mean that anything goes; the region of
acceptable parameters may be small compared to prior knowledge
about their values.  Yet it does validate a common approach
to modeling such systems, in which educated guesses are made
for most parameters, and a remaining handful are fit to the data.
In the common situation in which there are a small number $m$ of
important `stiff' directions, with
remaining sloppy directions extending to
cover the full range of feasible parameters,
fitting $m$ parameters will be enough to locate the sloppy
subspace. (And if using a
maximum likelihood approach, this is in fact
statistically preferred to fitting all parameters, in order
to avoid overfitting.)  Unfortunately, it is hard to know $m$
ahead of time, in general requiring a sampling scheme like MCMC
or a geodesic-following algorithm~\cite{TranstrumMS10,TranstrumMS11} to
ascertain the global structure of the sloppy subspace.

The problem of parameter estimation has been central to the field of
systems biology for many years.  The extremely large uncertainties in
parameter estimates led to the suggestion that accurate parameter
estimates might not be possible~\cite{GutenkunstWCBMS07}.  However, advances in
experimental design have suggested that such estimates might be
feasible after all~\cite{ApgarWWT10,VilelaVMVA09,ErgulerS11,TranstrumQ12},
although requiring considerable
experimental effort~\cite{ChachraTS11}.  The perspective provided by sloppy model
analysis provides at least two alternatives to this method of
operation.

First, in spite of the large number of parameters, complex biological
systems typically exhibit simple behavior that requires only a few
parameters to describe, analogous to how the diffusion equation can
describe microscopically diverse processes.  Attempting to accurately
infer all of the parameters in a complex biological
model~\cite{LeeSKHK03} is analogous to learning all of the mechanical
and electrical properties of water molecules in order to accurately
predict a diffusion constant.  It would involve considerable effort
(measuring {\em all} the microscopic parameters
accurately~\cite{GutenkunstWCBMS07}), while the diffusion constant
can be easily measured using collective experiments, and used to
determine the result of any other collective experiment.

Second, in many biological systems, there is considerable uncertainty
about the microscopic structure of the system.  Sloppiness suggests
that an effective model that is microscopically inaccurate
may still be insightful and
predictive in spite of getting many specific details wrong.  For
example a hopping model for thermal conductivity would be `wrong'
even though it gives the right thermal diffusion equation.  `Wrong'
models can provide key insights into the system level behavior
because they share important features with the true system.  In such a
scenario, it is the flexibility provided by
large uncertainties in the parameters that allows
the model to be useful.  Any attempt to infer all the microscopic
parameters would break the model, preventing it from being able to
fit the data.

Indeed, it is difficult to envision a completely microscopic model
in systems biology. Any model will have rates and binding affinities
that will be altered by the surrounding complex stew of proteins, ions,
lipids, and cellular substructures. Is the well-known dependence of a reaction
rate on salt concentration (described by an effective Gibbs free energy
tracing over the ionic degrees of freedom) qualitatively different
from the dependence of an effective reaction rate on cross-talk, 
regulatory mechanisms, or even parallel or competing pathways not incorporated
into the model? We are reminded of quantum field theories, where the properties
(say) of the electron known to quantum chemistry are {\em renormalized}
by electron-hole reactions in the surrounding vacuum which are ignored
and ignorable at low energies. Insofar as a model provides both insight
and correct predictions within its realm of validity, the fact that its
parameters have effective, renormalized values incorporating missing 
microscopic mechanisms should be expected, not disparaged.

\section{More general consequences of sloppiness}
\label{sec:Consequences}

The hyperribbon structures implied by interpolation theory
and information geometry in section~\ref{sec:WhySloppy} have profound
implications. Complex scientific models have predictions that vary
in far fewer ways than their complexity would indicate. Multiparameter
models have behavior that largely depend upon only a few combinations
of microscopic parameters. The high-dimensional results of a system with a 
large number of control parameters will be well encompassed by a rather
flat, low-dimensional manifold of behavior. In this section, we shall
speculate about these larger issues, and how they may explain the success
of our efforts to organize and understand our environment.

\noindent{\bf Efficacy of Principal Component Analysis.}
Principal component analysis, or PCA, has long been an effective tool
for data analysis. Given a high-dimensional data set, such as the changes
of mRNA levels for thousands of genes under several experimental
conditions~\cite{Ringner08}, PCA provides a 
reduced-dimensional description which often retains most of the variation
in the original data set in a few linear {\em components}. Arranging the
data into a matrix $R_{j n} + c_j$ of experiments
$n$ and data points $j$ centered at $c_j$, PCA uses 
the {\em singular value decomposition} (SVD)
\begin{align}
\label{eq:SVD}
R &= \sum_k \sigma_{k} \hat{\bu}^{(k)} \otimes \hat{\bv}^{(k)} \\
R_{j n} &= \sum_k \sigma_{k} \hat{u}^{(k)}_j \hat{v}^{(k)}_n
\end{align}
to write $R$ as the sum of outer products of orthonormal vectors
$\hat \bu^{(k)}$ in data space and $\hat \bv^{(k)}$ in the space
of experiments. Here
$\sigma_1 \ge \sigma_2 \ge ... \ge 0$ are nonnegative `strengths' of the
different components $k$.
These singular values can be viewed
as a generalization of eigenvalues for non-square, non-symmetric matrices.
The $\hat{\bu}^{(k)}$ for small $k$ describe the `long axes' of the
data, viewed as a cloud of points $\bR_n$ in data space; $\sigma_k$
is the RMS extent of the cloud in direction $\hat{\bu}^{(k)}$.
The utility of PCA stems from the fact that in many circumstances only
a few components $k$ are needed to provide an accurate reconstruction of the 
original data.  Just as our sloppy eigenvalues converge geometrically to 
zero, the singular values $\sigma_k$ often rapidly vanish. It is
straightforward to show that the truncated SVD
keeping only the first, largest $K$ components is an optimal approximation
to the data, with total least square error bounded by $\sum_{K+1} \sigma_k^2$.
These largest singular components often have physical or biological
interpretations -- sometimes mundane but useful (which machine was used
to take the data), sometimes scientifically central.

Why does Nature often demand so few components to describe large dimensional
data sets? Sloppiness provides a natural explanation. If the data
results from (say) a biological system whose behavior is described by a 
sloppy model $\by(\btheta)$, and if the different experiments are sampling
different parameter sets $\btheta_n$, then the data will be points
$R_{j n} + c_j = y_j(\btheta_n)$ on the model manifold. Insofar as the model
manifold has the hyperribbon structure we predict, it has only a few
long axes (corresponding to stiff directions)
and it is extrinsically very flat along these
axes~\cite[Fig.~18]{TranstrumMS11}. 
Here each $y_j - c_j$, being a difference between a data point and the center
of the data, will be nearly a linear sum of a small number $K$ of long
directions of the model manifold, and the RMS spread along this
$k^{\mathrm{th}}$ direction will be bounded by the width of the model
manifold in that direction, plus a small correction for the curvature.
As {\em any} cloud of experimental points must be bounded by the model
manifold, the high singular values will be bounded by the hierarchy of
widths of the hyperribbon.
Hence our arguments for the hyperribbon structure of the
model manifold in multiparameter models provide a fundamental
explanation for the success of PCA for these systems.

\noindent{\bf Efficacy of Levenberg-Marquardt; improved algorithms.}

The Levenberg-Marquardt
algorithm~\cite{Levenberg1944,Marquardt1963,Press2007} is one of the
standard algorithms for least squares minimization.  Its broad utility
can be explained through the lens of sloppy models and geometric
insights lead to natural improvements.  Minimizing a linear
approximation to a nonlinear model with a constraint on the step size
\begin{equation} 
\min_{\delta \theta} \left| y(\theta_0) + J \delta \theta - y_0 \right|^2, \ \ \ \left| \delta \theta \right| \leq \Delta
\end{equation}
leads to the iterative algorithm
\begin{equation}
\label{eq:LMstep}
\delta \theta = - \left( J^T J + \lambda \right)^{-1} J^T \left( y(\theta_0) - y_0 \right),
\end{equation}
where $\lambda$ is a Lagrange multiplier.  The FIM ($J^TJ$) for a
typical sloppy model is extremely ill-conditioned.  However, the
dampened scaling matrix $J^TJ + \lambda$ will have no eigenvalues
smaller than $\lambda$.  By tuning $\lambda$, the algorithm is able to
explicitly control the effects of sloppiness.  Furthermore, since the
eigenvalues of $J^T J$ are roughly log-linear, $\lambda$ need not be
finely tuned to be effective.  By slowly decreasing $\lambda$, the
algorithm fits the key features of the data first (i.e., the stiffest
directions), followed by successive refining approximations (i.e.,
progressively more sloppy components).  The algorithm may still
converge slowly as it navigates the extremely narrow canyons of the
cost surface (see Figure~\ref{fig:Ellipses}) or fail completely if it
becomes trapped near the boundary of the model
manifold~\cite{TranstrumMS10,TranstrumMS11}.

Information geometry provides a remarkable new perspective on the Levenberg
Marquardt algorithm. The move $\delta\theta$ for $\lambda=0$ is the direction
in parameter space corresponding to the steepest descent direction in data
space; for $\lambda\ne 0$ the move is the steepest descent direction on the
{\em model graph}~\cite{TranstrumMS10,TranstrumMS11}. The
fact that the model graph is extrinsically rather flat turns the 
narrow optimization valleys in parameter space into nearly concentric 
hyperspheres in data space -- explaining the power of the method. 
Levenberg-Marquardt takes steps along straight lines in parameter space;
to take full advantage of the flatness of the model manifold, it should
ideally move along geodesics. As it happens, the leading term in the 
geodesic equation is numerically cheap to calculate, providing a 
a ``geodesic acceleration'' correction to
the Levenberg-Marquardt algorithm which greatly improves the performance
and reliability of the
algorithm~\cite{TranstrumSnnb,url:GeodesicAccelerationSourceForge}.

\noindent{\bf Evolution is enabled.}
Besides practical consequences for parameter estimation of biochemical
networks (section~\ref{sec:biochemical}),
sloppiness has potential implications for biology and evolution.
Specifically, the fact that biological systems often achieve
remarkable robustness to environmental perturbations may be less
mysterious when taking into account the vastness of sloppy subspaces.
For instance, the circadian rhythm in cyanobacteria, controlled by the
dynamics of phosphorylation of three interacting Kai proteins, seems
remarkable in that it maintains a 24--hour cycle over a range of
temperature over which kinetic rates in the system are expected to
double.  Yet the degree of sloppiness in the system suggests that
evolution may have to tune only a few stiff parameter directions to
get the desired behavior at any given temperature, and perhaps only
one extra parameter direction to make that behavior robust to
temperature variation~\cite{DanielsCSGM08}.  Extended, high-dimensional 
neutral spaces have been identified as a central element underlying robustness 
and evolvability in living systems~\cite{Wagner05}, and sloppy parameter spaces 
play a similar role: a population with individuals spread
throughout a sloppy subspace can more easily reach a broader range of
phenotypic changes, such that the population is simultaneously highly
robust and highly evolvable~\cite{DanielsCSGM08}.

\noindent{\bf Pattern recognition as low-dimensional representation.} 
The pattern recognition methods we use to comprehend the world around us
are clearly low-dimensional representations. Cartoons embody this: we can
recognize and appreciate faces, motion, objects, and animals depicted with
a few pen strokes. In principle, one could distinguish different people
by patterns of scars, fingerprints, or retinal patterns, but our brains
instead process subtle overall features. Caricatures in particular build
on this low-dimensional representation -- exaggerating unusual features of 
the ears or nose of a celebrity makes them more recognizable, placing 
them farther along the relevant axes of some model manifold of facial features.
Archetypal analysis~\cite{Cutler94}, a branch of machine learning,
analyzes data sets with a matrix factorization similar to PCA, but
expressing data points as {\em convex} sums of features that are not constrained
to be orthogonal. In addition, the features must be convex combinations of 
data points. Archetypal analysis applied to suitably processed 
facial image data allows faces to be decomposed into strikingly meaningful
characteristic features~\cite{Cutler94,MorupH12,ThurauKB09}. 
The success of such algorithms is clearly related to a 
hidden low-dimensional representation of the data. One may speculate
that our facial structures are determined by the effects of genetic
and environmental control parameters $\btheta$, and that the resulting
model manifold of faces has a hyperribbon structure, explaining the
success of the linear, low-dimensional archetypal analysis methods, and
perhaps also the success of our biological pattern recognition skills.

\noindent{\bf Big data is reducible.} 
Machine learning methods that search for patterns in enormous data sets
are a growing feature of our information economy. These methods at root
discover low-dimensional representations of the high-dimensional data set.
Some tasks, such as the methods used to win the Netflix
challenge~\cite{KorenBV09} of predicting what movies users will like,
directly make use of this low-dimensional
representation by using SVD and PCA.
More complex problems, such as digital image recognition, make use
of artificial neural networks, such as stacked denoising
autoencoders~\cite{VincentLBM08}. Consider the problem of recognizing
handwritten digits (the MNIST database). 
The neural networks can be viewed
as a fitting problem, with parameters $\theta_\alpha$ giving the outputs
of the digital neurons, and the model $\by(\btheta)$ producing a digital
image that is optimized to best represent the written digits.
The training of these networks focuses on simultaneously developing a model 
manifold flexible enough to closely mimic the data set of digits, and
of developing a mapping $\tilde{\by}^{-1}(\bd)$ from the original data $\bd$
depicting the digit to neural outputs $\btheta = \tilde{\by}^{-1}(\bd)$
close to the best fit. Neural networks starting with high-dimensional data
routinely distill the data into a much smaller, more comprehensible set
of neural outputs $\btheta$ -- which are then used to classify or reconstruct
the original data. Initial explorations of a stacked denoising autoencoder
trained on the MNIST digit data by Hayden {\em et al.}~\cite{HaydenASnn}
show a clear
hyperribbon structure. What is surprising is not that the structure of a 
successful neural network has a hyperribbon structure. Indeed, if it were
not true that the $N+1$th thinnest direction on the model manifold is
significantly thinner than the first $N$ directions, surely an $N$ neuron
model would fail to capture the behavior of the data. What does demand
explanation is that these methods succeed at all -- that our handwritten
digits live on a hyperribbon, allowing the neural networks to succeed.

\noindent{\bf Science is possible.} 
In fields like ecology, systems biology, and macroeconomics, grossly 
simplified models capture important features of the behavior of incredibly
complex interacting systems. If what everyone ate for breakfast was crucial
in determining the economic productivity each day, and breakfast eating
habits were themselves not comprehensible,  macroeconomics would be 
doomed as a subject. We argue that adding more complexity to a model
produces diminishing returns in fidelity, {\em because the model predictions
have an underlying hyperribbon structure.}

\noindent{\bf Different models can describe the same behavior.}
We are told that science works by creating theories, and testing rival
theories with experiments to determine which is wrong. A more nuanced
view allows for effective theories of limited validity -- Newton wasn't wrong
and Einstein right, Newton's theory is valid when velocities are slow
compared to the speed of light. In more complex environments, several
theoretical descriptions can cast useful light onto the same phenomena
(`soft' and `hard' order parameters for magnets and liquid
crystals~\cite[Ch.~9]{Sethna06}). Also, in fields like economics and systems
biology, all descriptions are doomed to neglect pathways or behavior without
the justification of a small parameter. So long as these models are capable
of capturing the `long axes' of the model manifold in the data space of
known behavior, and are successful at predicting the behavior in the larger
data space of experiments of interest, one must view them as successful.
Many such models will in general exist -- certainly reduced models extracted
systematically from a microscopic model (section~\ref{sec:ModelReduction}),
but other models as well. Naturally, one should design experiments that
test the limits of these models, and cleanly discriminate between rival models.
Our information geometry methods could be useful in the design of experiments
distinguishing rival models; current methods that linearize about expected
behavior~\cite{CaseyBFGWMBCS07} could be replaced by geometric
methods that allow for large uncertainties in model parameters 
corresponding to nearly indistinguishable model predictions.

\noindent{\bf Why is the world comprehensible?} 
Surely the reason that handwritten digits have a hyperribbon structure --
that we don't use random dot patterns to write numbers -- is partially
related to the way our brain is wired. We recognize cartoons easily, therefore
the information in our handwriting is encapsulated in cartoon-like 
subrepresentations. Surely physics has low-dimensional representations
(section~\ref{sec:Physics})
independently of the way our brain works. The continuum limit describes
our world perturbatively in the inverse length and time scales of the
observation; the renormalization group in addition perturbs in the 
distance to the critical point. Why is science successful
in other fields, systems biology and macroeconomics, for example?
Is it a selection effect -- do we choose to study subjects where our
brains see patterns (low-dimensional representations), and then describe
those patterns using theories with hyperribbon structures? Or are there
deep underpinning structures (evolution, game theory) that guide the
behavior into comprehensible patterns? A cellular control circuit
where hundreds of parameters all individually control important, different 
aspects of the behavior would be incomprehensible without full microscopic
information, discouraging us from trying to model it. On the other hand,
it would seem challenging for such a circuit to arise under Darwinian
evolution. Perhaps modularity and comprehensibility 
themselves are the result of evolution~\cite{KirschnerG98,HartwellHLM99,KashtanA05,CluneML13}.

\noindent{\bf Conclusion.}
What began as a rather pragmatic exercise in parameter fitting has
blossomed into an enterprise that stretches across the landscape of
science.  The work described here has both methodological implications
for the development and validation of scientific models (in the areas
of optimization, machine learning and model reduction) as well as
philosophical implications for how we reason about the world around
us.  By investigating and characterizing in detail the geometric and
topological structures underlying scientific models, this work
connects bottom-up descriptions of complex processes with top-down
inferences drawn from data, paving the way for emergent theories in
physics, biology, and beyond.

\begin{acknowledgments}
We would like to thank Alex Alemi, Phil Burnham, Colin Clement, Josh Fass, Ryan Gutenkunst, Lorien Hayden, Lei Huang, Jaron Kent-Dobias, Ben Nicholson and Hao Shi for their assistance
and insights. This work was supported in part by NSF DMR 1312160 (JPS), NSF IOS 1127017 (CRM), the John Templeton Foundation through a grant to SFI to study complexity (BCD), the U.S.~Army Research Laboratory and the U.S.~Army Research Office under contract number W911NF-13-1-0340 (BCD).
\end{acknowledgments}

\include{Bibliography}

\end{document}

%% file: JPSCommandsJCP.tex
\usepackage{bbm}
\usepackage{dsfont}

\newcommand{\JPSfrac}[2]{\ensuremath{{}^{#1}\mspace{-6.5mu}/\mspace{-4.5mu}_{#2}}}

\newcommand{\half}{{\JPSfrac{1}{2}}}

\newcommand{\Rbb}{{\mathbb R}}

\newcommand{\btheta}{{\mbox{\boldmath$\theta$}}}

\newcommand{\bd}{{\mathbf{d}}}

\newcommand{\bR}{{\mathbf{R}}}

\newcommand{\bu}{{\mathbf{u}}}

\newcommand{\bv}{{\mathbf{v}}}

\newcommand{\by}{{\mathbf{y}}}



%% file: Bibliography.tex
\bibliography{SethnaRecs,refs}

%% file: JChemPhysReview.bbl
\begin{thebibliography}{78}%
\makeatletter
\providecommand \@ifxundefined [1]{%
 \@ifx{#1\undefined}
}%
\providecommand \@ifnum [1]{%
 \ifnum #1\expandafter \@firstoftwo
 \else \expandafter \@secondoftwo
 \fi
}%
\providecommand \@ifx [1]{%
 \ifx #1\expandafter \@firstoftwo
 \else \expandafter \@secondoftwo
 \fi
}%
\providecommand \natexlab [1]{#1}%
\providecommand \enquote  [1]{``#1''}%
\providecommand \bibnamefont  [1]{#1}%
\providecommand \bibfnamefont [1]{#1}%
\providecommand \citenamefont [1]{#1}%
\providecommand \href@noop [0]{\@secondoftwo}%
\providecommand \href [0]{\begingroup \@sanitize@url \@href}%
\providecommand \@href[1]{\@@startlink{#1}\@@href}%
\providecommand \@@href[1]{\endgroup#1\@@endlink}%
\providecommand \@sanitize@url [0]{\catcode `\\12\catcode `\$12\catcode
  `\&12\catcode `\#12\catcode `\^12\catcode `\_12\catcode `\%12\relax}%
\providecommand \@@startlink[1]{}%
\providecommand \@@endlink[0]{}%
\providecommand \url  [0]{\begingroup\@sanitize@url \@url }%
\providecommand \@url [1]{\endgroup\@href {#1}{\urlprefix }}%
\providecommand \urlprefix  [0]{URL }%
\providecommand \Eprint [0]{\href }%
\providecommand \doibase [0]{http://dx.doi.org/}%
\providecommand \selectlanguage [0]{\@gobble}%
\providecommand \bibinfo  [0]{\@secondoftwo}%
\providecommand \bibfield  [0]{\@secondoftwo}%
\providecommand \translation [1]{[#1]}%
\providecommand \BibitemOpen [0]{}%
\providecommand \bibitemStop [0]{}%
\providecommand \bibitemNoStop [0]{.\EOS\space}%
\providecommand \EOS [0]{\spacefactor3000\relax}%
\providecommand \BibitemShut  [1]{\csname bibitem#1\endcsname}%
\let\auto@bib@innerbib\@empty
\bibitem [{\citenamefont {Dyson}(2004)}]{Dyson:2004}%
  \BibitemOpen
  \bibfield  {author} {\bibinfo {author} {\bibfnamefont {F.}~\bibnamefont
  {Dyson}},\ }\href@noop {} {\bibfield  {journal} {\bibinfo  {journal}
  {Nature}\ }\textbf {\bibinfo {volume} {427}},\ \bibinfo {pages} {297}
  (\bibinfo {year} {2004})}\BibitemShut {NoStop}%
\bibitem [{\citenamefont {Ditley}, \citenamefont {Mayer},\ and\ \citenamefont
  {Loew}(2013)}]{Loew13}%
  \BibitemOpen
  \bibfield  {author} {\bibinfo {author} {\bibfnamefont {J.}~\bibnamefont
  {Ditley}}, \bibinfo {author} {\bibfnamefont {B.}~\bibnamefont {Mayer}}, \
  and\ \bibinfo {author} {\bibfnamefont {L.}~\bibnamefont {Loew}},\ }\href@noop
  {} {\bibfield  {journal} {\bibinfo  {journal} {Biophysical Journal}\ }\textbf
  {\bibinfo {volume} {104}},\ \bibinfo {pages} {520} (\bibinfo {year}
  {2013})}\BibitemShut {NoStop}%
\bibitem [{\citenamefont {Brown}\ and\ \citenamefont
  {Sethna}(2003)}]{BrownS03}%
  \BibitemOpen
  \bibfield  {author} {\bibinfo {author} {\bibfnamefont {K.~S.}\ \bibnamefont
  {Brown}}\ and\ \bibinfo {author} {\bibfnamefont {J.~P.}\ \bibnamefont
  {Sethna}},\ }\href@noop {} {\bibfield  {journal} {\bibinfo  {journal}
  {Physical Review E}\ }\textbf {\bibinfo {volume} {68}},\ \bibinfo {pages}
  {021904} (\bibinfo {year} {2003})}\BibitemShut {NoStop}%
\bibitem [{\citenamefont {Brown}\ \emph {et~al.}(2004)\citenamefont {Brown},
  \citenamefont {Hill}, \citenamefont {Calero}, \citenamefont {Myers},
  \citenamefont {Lee}, \citenamefont {Sethna},\ and\ \citenamefont
  {Cerione}}]{BrownHCMLSC04}%
  \BibitemOpen
  \bibfield  {author} {\bibinfo {author} {\bibfnamefont {K.~S.}\ \bibnamefont
  {Brown}}, \bibinfo {author} {\bibfnamefont {C.~C.}\ \bibnamefont {Hill}},
  \bibinfo {author} {\bibfnamefont {G.~A.}\ \bibnamefont {Calero}}, \bibinfo
  {author} {\bibfnamefont {C.~R.}\ \bibnamefont {Myers}}, \bibinfo {author}
  {\bibfnamefont {K.~H.}\ \bibnamefont {Lee}}, \bibinfo {author} {\bibfnamefont
  {J.~P.}\ \bibnamefont {Sethna}}, \ and\ \bibinfo {author} {\bibfnamefont
  {R.~A.}\ \bibnamefont {Cerione}},\ }\href@noop {} {\bibfield  {journal}
  {\bibinfo  {journal} {Physical Biology}\ }\textbf {\bibinfo {volume} {1}},\
  \bibinfo {pages} {184} (\bibinfo {year} {2004})}\BibitemShut {NoStop}%
\bibitem [{\citenamefont {Waterfall}\ \emph {et~al.}(2006)\citenamefont
  {Waterfall}, \citenamefont {Casey}, \citenamefont {Gutenkunst}, \citenamefont
  {Brown}, \citenamefont {Myers}, \citenamefont {Brouwer}, \citenamefont
  {Elser},\ and\ \citenamefont {Sethna}}]{WaterfallCGBMBES06}%
  \BibitemOpen
  \bibfield  {author} {\bibinfo {author} {\bibfnamefont {J.~J.}\ \bibnamefont
  {Waterfall}}, \bibinfo {author} {\bibfnamefont {F.~P.}\ \bibnamefont
  {Casey}}, \bibinfo {author} {\bibfnamefont {R.~N.}\ \bibnamefont
  {Gutenkunst}}, \bibinfo {author} {\bibfnamefont {K.~S.}\ \bibnamefont
  {Brown}}, \bibinfo {author} {\bibfnamefont {C.~R.}\ \bibnamefont {Myers}},
  \bibinfo {author} {\bibfnamefont {P.~W.}\ \bibnamefont {Brouwer}}, \bibinfo
  {author} {\bibfnamefont {V.}~\bibnamefont {Elser}}, \ and\ \bibinfo {author}
  {\bibfnamefont {J.~P.}\ \bibnamefont {Sethna}},\ }\href@noop {} {\bibfield
  {journal} {\bibinfo  {journal} {Physical Review Letters}\ }\textbf {\bibinfo
  {volume} {97}},\ \bibinfo {pages} {150601} (\bibinfo {year}
  {2006})}\BibitemShut {NoStop}%
\bibitem [{\citenamefont {Frederiksen}\ \emph {et~al.}(2004)\citenamefont
  {Frederiksen}, \citenamefont {Jacobsen}, \citenamefont {Brown},\ and\
  \citenamefont {Sethna}}]{FrederiksenJBS04}%
  \BibitemOpen
  \bibfield  {author} {\bibinfo {author} {\bibfnamefont {S.~L.}\ \bibnamefont
  {Frederiksen}}, \bibinfo {author} {\bibfnamefont {K.~W.}\ \bibnamefont
  {Jacobsen}}, \bibinfo {author} {\bibfnamefont {K.~S.}\ \bibnamefont {Brown}},
  \ and\ \bibinfo {author} {\bibfnamefont {J.~P.}\ \bibnamefont {Sethna}},\
  }\href@noop {} {\bibfield  {journal} {\bibinfo  {journal} {Physical Review
  Letters}\ }\textbf {\bibinfo {volume} {93}},\ \bibinfo {pages} {216401}
  (\bibinfo {year} {2004})}\BibitemShut {NoStop}%
\bibitem [{\citenamefont {Gutenkunst}(2007)}]{GutenkunstPhD}%
  \BibitemOpen
  \bibfield  {author} {\bibinfo {author} {\bibfnamefont {R.}~\bibnamefont
  {Gutenkunst}},\ }\emph {\bibinfo {title} {Sloppiness, Modeling, and Evolution
  in Biochemical Networks}},\ \href@noop {} {Ph.D. thesis},\ \bibinfo  {school}
  {Cornell University} (\bibinfo {year} {2007}),\ \bibinfo {note}
  {\url{http://ecommons.library.cornell.edu/handle/1813/8206}}\BibitemShut
  {NoStop}%
\bibitem [{\citenamefont {BERMAN}\ and\ \citenamefont
  {WANG}(2007)}]{BermanW07}%
  \BibitemOpen
  \bibfield  {author} {\bibinfo {author} {\bibfnamefont {G.~J.}\ \bibnamefont
  {BERMAN}}\ and\ \bibinfo {author} {\bibfnamefont {Z.~J.}\ \bibnamefont
  {WANG}},\ }\href {\doibase 10.1017/S0022112007006209} {\bibfield  {journal}
  {\bibinfo  {journal} {Journal of Fluid Mechanics}\ }\textbf {\bibinfo
  {volume} {582}},\ \bibinfo {pages} {153} (\bibinfo {year}
  {2007})}\BibitemShut {NoStop}%
\bibitem [{\citenamefont {Machta}\ \emph {et~al.}(2013)\citenamefont {Machta},
  \citenamefont {Chachra}, \citenamefont {Transtrum},\ and\ \citenamefont
  {Sethna}}]{MachtaCTS13}%
  \BibitemOpen
  \bibfield  {author} {\bibinfo {author} {\bibfnamefont {B.~B.}\ \bibnamefont
  {Machta}}, \bibinfo {author} {\bibfnamefont {R.}~\bibnamefont {Chachra}},
  \bibinfo {author} {\bibfnamefont {M.}~\bibnamefont {Transtrum}}, \ and\
  \bibinfo {author} {\bibfnamefont {J.~P.}\ \bibnamefont {Sethna}},\
  }\href@noop {} {\bibfield  {journal} {\bibinfo  {journal} {Science}\ }\textbf
  {\bibinfo {volume} {342}},\ \bibinfo {pages} {604} (\bibinfo {year}
  {2013})}\BibitemShut {NoStop}%
\bibitem [{\citenamefont {Ruhe}(1980)}]{ruhe1980fitting}%
  \BibitemOpen
  \bibfield  {author} {\bibinfo {author} {\bibfnamefont {A.}~\bibnamefont
  {Ruhe}},\ }\href@noop {} {\bibfield  {journal} {\bibinfo  {journal} {SIAM
  Journal on Scientific and Statistical Computing}\ }\textbf {\bibinfo {volume}
  {1}},\ \bibinfo {pages} {481} (\bibinfo {year} {1980})}\BibitemShut {NoStop}%
\bibitem [{\citenamefont {Transtrum}, \citenamefont {Machta},\ and\
  \citenamefont {Sethna}(2011)}]{TranstrumMS11}%
  \BibitemOpen
  \bibfield  {author} {\bibinfo {author} {\bibfnamefont {M.~K.}\ \bibnamefont
  {Transtrum}}, \bibinfo {author} {\bibfnamefont {B.~B.}\ \bibnamefont
  {Machta}}, \ and\ \bibinfo {author} {\bibfnamefont {J.~P.}\ \bibnamefont
  {Sethna}},\ }\href@noop {} {\bibfield  {journal} {\bibinfo  {journal} {Phys.
  Rev. E}\ }\textbf {\bibinfo {volume} {83}},\ \bibinfo {pages} {036701}
  (\bibinfo {year} {2011})}\BibitemShut {NoStop}%
\bibitem [{\citenamefont {Gutenkunst}\ \emph {et~al.}(2007)\citenamefont
  {Gutenkunst}, \citenamefont {Waterfall}, \citenamefont {Casey}, \citenamefont
  {Brown}, \citenamefont {Myers},\ and\ \citenamefont
  {Sethna}}]{GutenkunstWCBMS07}%
  \BibitemOpen
  \bibfield  {author} {\bibinfo {author} {\bibfnamefont {R.~N.}\ \bibnamefont
  {Gutenkunst}}, \bibinfo {author} {\bibfnamefont {J.~J.}\ \bibnamefont
  {Waterfall}}, \bibinfo {author} {\bibfnamefont {F.~P.}\ \bibnamefont
  {Casey}}, \bibinfo {author} {\bibfnamefont {K.~S.}\ \bibnamefont {Brown}},
  \bibinfo {author} {\bibfnamefont {C.~R.}\ \bibnamefont {Myers}}, \ and\
  \bibinfo {author} {\bibfnamefont {J.~P.}\ \bibnamefont {Sethna}},\
  }\href@noop {} {\bibfield  {journal} {\bibinfo  {journal} {P{L}o{S}
  Computational Biology}\ }\textbf {\bibinfo {volume} {3}},\ \bibinfo {pages}
  {1871} (\bibinfo {year} {2007})}\BibitemShut {NoStop}%
\bibitem [{\citenamefont {Wigner}(1960)}]{Wigner60}%
  \BibitemOpen
  \bibfield  {author} {\bibinfo {author} {\bibfnamefont {E.~P.}\ \bibnamefont
  {Wigner}},\ }\href@noop {} {\bibfield  {journal} {\bibinfo  {journal}
  {Communications on Pure and Applied Mathematics}\ }\textbf {\bibinfo {volume}
  {13}},\ \bibinfo {pages} {1} (\bibinfo {year} {1960})}\BibitemShut {NoStop}%
\bibitem [{\citenamefont {Anderson}\ \emph {et~al.}(1972)\citenamefont
  {Anderson} \emph {et~al.}}]{anderson1972more}%
  \BibitemOpen
  \bibfield  {author} {\bibinfo {author} {\bibfnamefont {P.~W.}\ \bibnamefont
  {Anderson}} \emph {et~al.},\ }\href@noop {} {\bibfield  {journal} {\bibinfo
  {journal} {Science}\ }\textbf {\bibinfo {volume} {177}},\ \bibinfo {pages}
  {393} (\bibinfo {year} {1972})}\BibitemShut {NoStop}%
\bibitem [{\citenamefont {Amari}\ and\ \citenamefont
  {Nagaoka}(2000)}]{Amari00}%
  \BibitemOpen
  \bibfield  {author} {\bibinfo {author} {\bibfnamefont {S.}~\bibnamefont
  {Amari}}\ and\ \bibinfo {author} {\bibfnamefont {H.}~\bibnamefont
  {Nagaoka}},\ }\href {http://books.google.com/books?id=XHGo0sEvr-AC} {\emph
  {\bibinfo {title} {Methods of Information Geometry}}},\ Translations of
  Mathematical Monographs\ (\bibinfo  {publisher} {American Mathematical
  Society},\ \bibinfo {year} {2000})\BibitemShut {NoStop}%
\bibitem [{\citenamefont {Averick}\ \emph {et~al.}(1992)\citenamefont
  {Averick}, \citenamefont {Carter}, \citenamefont {More},\ and\ \citenamefont
  {Xue}}]{Averick1992}%
  \BibitemOpen
  \bibfield  {author} {\bibinfo {author} {\bibfnamefont {B.}~\bibnamefont
  {Averick}}, \bibinfo {author} {\bibfnamefont {R.}~\bibnamefont {Carter}},
  \bibinfo {author} {\bibfnamefont {J.}~\bibnamefont {More}}, \ and\ \bibinfo
  {author} {\bibfnamefont {G.}~\bibnamefont {Xue}},\ }\href@noop {} {\bibfield
  {journal} {\bibinfo  {journal} {Preprint MCS-P153-0694, Mathematics and
  Computer Science Division, Argonne National Laboratory, Argonne, Illinois}\ }
  (\bibinfo {year} {1992})}\BibitemShut {NoStop}%
\bibitem [{\citenamefont {Kowalik}\ and\ \citenamefont
  {Morrison}(1968)}]{kowalik1968analysis}%
  \BibitemOpen
  \bibfield  {author} {\bibinfo {author} {\bibfnamefont {J.}~\bibnamefont
  {Kowalik}}\ and\ \bibinfo {author} {\bibfnamefont {J.}~\bibnamefont
  {Morrison}},\ }\href@noop {} {\bibfield  {journal} {\bibinfo  {journal}
  {Mathematical Biosciences}\ }\textbf {\bibinfo {volume} {2}},\ \bibinfo
  {pages} {57} (\bibinfo {year} {1968})}\BibitemShut {NoStop}%
\bibitem [{\citenamefont {Beale}(1960)}]{beale1960confidence}%
  \BibitemOpen
  \bibfield  {author} {\bibinfo {author} {\bibfnamefont {E.}~\bibnamefont
  {Beale}},\ }\href@noop {} {\bibfield  {journal} {\bibinfo  {journal} {Journal
  of the Royal Statistical Society. Series B (Methodological)}\ ,\ \bibinfo
  {pages} {41}} (\bibinfo {year} {1960})}\BibitemShut {NoStop}%
\bibitem [{\citenamefont {Bates}\ and\ \citenamefont
  {Watts}(1980)}]{bates1980relative}%
  \BibitemOpen
  \bibfield  {author} {\bibinfo {author} {\bibfnamefont {D.~M.}\ \bibnamefont
  {Bates}}\ and\ \bibinfo {author} {\bibfnamefont {D.~G.}\ \bibnamefont
  {Watts}},\ }\href@noop {} {\bibfield  {journal} {\bibinfo  {journal} {Journal
  of the Royal Statistical Society. Series B (Methodological)}\ ,\ \bibinfo
  {pages} {1}} (\bibinfo {year} {1980})}\BibitemShut {NoStop}%
\bibitem [{\citenamefont {Amari}(1985)}]{amari1985differential}%
  \BibitemOpen
  \bibfield  {author} {\bibinfo {author} {\bibfnamefont {S.-i.}\ \bibnamefont
  {Amari}},\ }\href@noop {} {\emph {\bibinfo {title} {Differential-geometrical
  methods in statistics}}}\ (\bibinfo  {publisher} {Springer},\ \bibinfo {year}
  {1985})\BibitemShut {NoStop}%
\bibitem [{\citenamefont {Amari}\ \emph {et~al.}(1987)\citenamefont {Amari},
  \citenamefont {Barndorff-Nielsen}, \citenamefont {Kass}, \citenamefont
  {Lauritzen},\ and\ \citenamefont {Rao}}]{amari1987differential}%
  \BibitemOpen
  \bibfield  {author} {\bibinfo {author} {\bibfnamefont {S.-I.}\ \bibnamefont
  {Amari}}, \bibinfo {author} {\bibfnamefont {O.~E.}\ \bibnamefont
  {Barndorff-Nielsen}}, \bibinfo {author} {\bibfnamefont {R.}~\bibnamefont
  {Kass}}, \bibinfo {author} {\bibfnamefont {S.}~\bibnamefont {Lauritzen}}, \
  and\ \bibinfo {author} {\bibfnamefont {C.}~\bibnamefont {Rao}},\ }\href@noop
  {} {\bibfield  {journal} {\bibinfo  {journal} {Lecture Notes-Monograph
  Series}\ ,\ \bibinfo {pages} {i}} (\bibinfo {year} {1987})}\BibitemShut
  {NoStop}%
\bibitem [{\citenamefont {Murray}\ and\ \citenamefont
  {Rice}(1993)}]{murray1993differential}%
  \BibitemOpen
  \bibfield  {author} {\bibinfo {author} {\bibfnamefont {M.~K.}\ \bibnamefont
  {Murray}}\ and\ \bibinfo {author} {\bibfnamefont {J.~W.}\ \bibnamefont
  {Rice}},\ }\href@noop {} {\emph {\bibinfo {title} {Differential geometry and
  statistics}}},\ Vol.~\bibinfo {volume} {48}\ (\bibinfo  {publisher} {CRC
  Press},\ \bibinfo {year} {1993})\BibitemShut {NoStop}%
\bibitem [{\citenamefont {Amari}\ and\ \citenamefont
  {Nagaoka}(2007)}]{amari2007methods}%
  \BibitemOpen
  \bibfield  {author} {\bibinfo {author} {\bibfnamefont {S.-i.}\ \bibnamefont
  {Amari}}\ and\ \bibinfo {author} {\bibfnamefont {H.}~\bibnamefont
  {Nagaoka}},\ }\href@noop {} {\emph {\bibinfo {title} {Methods of information
  geometry}}},\ Vol.\ \bibinfo {volume} {191}\ (\bibinfo  {publisher} {American
  Mathematical Soc.},\ \bibinfo {year} {2007})\BibitemShut {NoStop}%
\bibitem [{\citenamefont {Transtrum}, \citenamefont {Machta},\ and\
  \citenamefont {Sethna}(2010)}]{TranstrumMS10}%
  \BibitemOpen
  \bibfield  {author} {\bibinfo {author} {\bibfnamefont {M.~K.}\ \bibnamefont
  {Transtrum}}, \bibinfo {author} {\bibfnamefont {B.~B.}\ \bibnamefont
  {Machta}}, \ and\ \bibinfo {author} {\bibfnamefont {J.~P.}\ \bibnamefont
  {Sethna}},\ }\href@noop {} {\bibfield  {journal} {\bibinfo  {journal} {Phys.
  Rev. Lett.}\ }\textbf {\bibinfo {volume} {104}},\ \bibinfo {pages} {060201}
  (\bibinfo {year} {2010})}\BibitemShut {NoStop}%
\bibitem [{\citenamefont {Spivak}(1979)}]{spivak1979}%
  \BibitemOpen
  \bibfield  {author} {\bibinfo {author} {\bibfnamefont {M.}~\bibnamefont
  {Spivak}},\ }\href@noop {} {\emph {\bibinfo {title} {A comprehensive
  introduction to differential geometry}}}\ (\bibinfo  {publisher} {Publish or
  Perish},\ \bibinfo {year} {1979})\BibitemShut {NoStop}%
\bibitem [{\citenamefont {Ivancevic}(2007)}]{Ivancevic2007}%
  \BibitemOpen
  \bibfield  {author} {\bibinfo {author} {\bibfnamefont {T.}~\bibnamefont
  {Ivancevic}},\ }\href@noop {} {\emph {\bibinfo {title} {Applied differential
  geometry: a modern introduction}}}\ (\bibinfo  {publisher} {World Scientific
  Pub Co Inc},\ \bibinfo {year} {2007})\BibitemShut {NoStop}%
\bibitem [{\citenamefont {Bates}\ and\ \citenamefont
  {Watts}(1981)}]{Bates1981}%
  \BibitemOpen
  \bibfield  {author} {\bibinfo {author} {\bibfnamefont {D.}~\bibnamefont
  {Bates}}\ and\ \bibinfo {author} {\bibfnamefont {D.}~\bibnamefont {Watts}},\
  }\href@noop {} {\bibfield  {journal} {\bibinfo  {journal} {Ann. Statist}\
  }\textbf {\bibinfo {volume} {9}},\ \bibinfo {pages} {1152} (\bibinfo {year}
  {1981})}\BibitemShut {NoStop}%
\bibitem [{\citenamefont {Bates}, \citenamefont {Hamilton},\ and\ \citenamefont
  {Watts}(1983)}]{Bates1983}%
  \BibitemOpen
  \bibfield  {author} {\bibinfo {author} {\bibfnamefont {D.}~\bibnamefont
  {Bates}}, \bibinfo {author} {\bibfnamefont {D.}~\bibnamefont {Hamilton}}, \
  and\ \bibinfo {author} {\bibfnamefont {D.}~\bibnamefont {Watts}},\
  }\href@noop {} {\bibfield  {journal} {\bibinfo  {journal} {Communications in
  Statistics-Simulation and Computation}\ }\textbf {\bibinfo {volume} {12}},\
  \bibinfo {pages} {469} (\bibinfo {year} {1983})}\BibitemShut {NoStop}%
\bibitem [{\citenamefont {Bates}\ and\ \citenamefont
  {Watts}(1988)}]{Bates1988}%
  \BibitemOpen
  \bibfield  {author} {\bibinfo {author} {\bibfnamefont {D.}~\bibnamefont
  {Bates}}\ and\ \bibinfo {author} {\bibfnamefont {D.}~\bibnamefont {Watts}},\
  }\href@noop {} {\emph {\bibinfo {title} {Nonlinear Regression Analysis and
  Its Applications}}}\ (\bibinfo  {publisher} {John Wiley},\ \bibinfo {year}
  {1988})\BibitemShut {NoStop}%
\bibitem [{\citenamefont {Wei}\ and\ \citenamefont
  {Kuo}(1969)}]{wei1969lumping}%
  \BibitemOpen
  \bibfield  {author} {\bibinfo {author} {\bibfnamefont {J.}~\bibnamefont
  {Wei}}\ and\ \bibinfo {author} {\bibfnamefont {J.~C.}\ \bibnamefont {Kuo}},\
  }\href@noop {} {\bibfield  {journal} {\bibinfo  {journal} {Industrial \&
  Engineering chemistry fundamentals}\ }\textbf {\bibinfo {volume} {8}},\
  \bibinfo {pages} {114} (\bibinfo {year} {1969})}\BibitemShut {NoStop}%
\bibitem [{\citenamefont {Liao}\ and\ \citenamefont
  {Lightfoot}(1988)}]{liao1988lumping}%
  \BibitemOpen
  \bibfield  {author} {\bibinfo {author} {\bibfnamefont {J.~C.}\ \bibnamefont
  {Liao}}\ and\ \bibinfo {author} {\bibfnamefont {E.~N.}\ \bibnamefont
  {Lightfoot}},\ }\href@noop {} {\bibfield  {journal} {\bibinfo  {journal}
  {Biotechnology and bioengineering}\ }\textbf {\bibinfo {volume} {31}},\
  \bibinfo {pages} {869} (\bibinfo {year} {1988})}\BibitemShut {NoStop}%
\bibitem [{\citenamefont {Huang}\ \emph {et~al.}(2005)\citenamefont {Huang},
  \citenamefont {Fairweather}, \citenamefont {Griffiths}, \citenamefont
  {Tomlin},\ and\ \citenamefont {Brad}}]{huang2005systematic}%
  \BibitemOpen
  \bibfield  {author} {\bibinfo {author} {\bibfnamefont {H.}~\bibnamefont
  {Huang}}, \bibinfo {author} {\bibfnamefont {M.}~\bibnamefont {Fairweather}},
  \bibinfo {author} {\bibfnamefont {J.}~\bibnamefont {Griffiths}}, \bibinfo
  {author} {\bibfnamefont {A.}~\bibnamefont {Tomlin}}, \ and\ \bibinfo {author}
  {\bibfnamefont {R.}~\bibnamefont {Brad}},\ }\href@noop {} {\bibfield
  {journal} {\bibinfo  {journal} {Proceedings of the Combustion Institute}\
  }\textbf {\bibinfo {volume} {30}},\ \bibinfo {pages} {1309} (\bibinfo {year}
  {2005})}\BibitemShut {NoStop}%
\bibitem [{\citenamefont {Goldenfeld}(1992)}]{goldenfeld1992lectures}%
  \BibitemOpen
  \bibfield  {author} {\bibinfo {author} {\bibfnamefont {N.}~\bibnamefont
  {Goldenfeld}},\ }\href@noop {} {\emph {\bibinfo {title} {Lectures on phase
  transitions and the renormalization group}}}\ (\bibinfo  {publisher}
  {Addison-Wesley, Advanced Book Program, Reading},\ \bibinfo {year}
  {1992})\BibitemShut {NoStop}%
\bibitem [{\citenamefont {Zinn-Justin}(2007)}]{zinn2007phase}%
  \BibitemOpen
  \bibfield  {author} {\bibinfo {author} {\bibfnamefont {J.}~\bibnamefont
  {Zinn-Justin}},\ }\href@noop {} {\emph {\bibinfo {title} {Phase transitions
  and renormalization group}}}\ (\bibinfo  {publisher} {Oxford University
  Press},\ \bibinfo {year} {2007})\BibitemShut {NoStop}%
\bibitem [{\citenamefont {Saksena}, \citenamefont {O'reilly},\ and\
  \citenamefont {Kokotovic}(1984)}]{saksena1984singular}%
  \BibitemOpen
  \bibfield  {author} {\bibinfo {author} {\bibfnamefont {V.}~\bibnamefont
  {Saksena}}, \bibinfo {author} {\bibfnamefont {J.}~\bibnamefont {O'reilly}}, \
  and\ \bibinfo {author} {\bibfnamefont {P.~V.}\ \bibnamefont {Kokotovic}},\
  }\href@noop {} {\bibfield  {journal} {\bibinfo  {journal} {Automatica}\
  }\textbf {\bibinfo {volume} {20}},\ \bibinfo {pages} {273} (\bibinfo {year}
  {1984})}\BibitemShut {NoStop}%
\bibitem [{\citenamefont {Kokotovic}, \citenamefont {Khali},\ and\
  \citenamefont {O'Reilly}(1999)}]{kokotovic1999singular}%
  \BibitemOpen
  \bibfield  {author} {\bibinfo {author} {\bibfnamefont {P.}~\bibnamefont
  {Kokotovic}}, \bibinfo {author} {\bibfnamefont {H.~K.}\ \bibnamefont
  {Khali}}, \ and\ \bibinfo {author} {\bibfnamefont {J.}~\bibnamefont
  {O'Reilly}},\ }\href@noop {} {\emph {\bibinfo {title} {Singular perturbation
  methods in control: analysis and design}}},\ Vol.~\bibinfo {volume} {25}\
  (\bibinfo  {publisher} {Siam},\ \bibinfo {year} {1999})\BibitemShut {NoStop}%
\bibitem [{\citenamefont {Naidu}(2002)}]{naidu2002singular}%
  \BibitemOpen
  \bibfield  {author} {\bibinfo {author} {\bibfnamefont {D.}~\bibnamefont
  {Naidu}},\ }\href@noop {} {\bibfield  {journal} {\bibinfo  {journal}
  {Dynamics of Continuous Discrete and Impulsive Systems Series B}\ }\textbf
  {\bibinfo {volume} {9}},\ \bibinfo {pages} {233} (\bibinfo {year}
  {2002})}\BibitemShut {NoStop}%
\bibitem [{\citenamefont {Antoulas}(2005)}]{antoulas2005approximation}%
  \BibitemOpen
  \bibfield  {author} {\bibinfo {author} {\bibfnamefont {A.~C.}\ \bibnamefont
  {Antoulas}},\ }\href@noop {} {\emph {\bibinfo {title} {Approximation of
  large-scale dynamical systems}}},\ Vol.~\bibinfo {volume} {6}\ (\bibinfo
  {publisher} {Siam},\ \bibinfo {year} {2005})\BibitemShut {NoStop}%
\bibitem [{\citenamefont {Lee}\ and\ \citenamefont
  {Othmer}(2010)}]{lee2010multi}%
  \BibitemOpen
  \bibfield  {author} {\bibinfo {author} {\bibfnamefont {C.~H.}\ \bibnamefont
  {Lee}}\ and\ \bibinfo {author} {\bibfnamefont {H.~G.}\ \bibnamefont
  {Othmer}},\ }\href@noop {} {\bibfield  {journal} {\bibinfo  {journal}
  {Journal of mathematical biology}\ }\textbf {\bibinfo {volume} {60}},\
  \bibinfo {pages} {387} (\bibinfo {year} {2010})}\BibitemShut {NoStop}%
\bibitem [{\citenamefont {Moore}(1981)}]{moore1981principal}%
  \BibitemOpen
  \bibfield  {author} {\bibinfo {author} {\bibfnamefont {B.}~\bibnamefont
  {Moore}},\ }\href@noop {} {\bibfield  {journal} {\bibinfo  {journal}
  {Automatic Control, IEEE Transactions on}\ }\textbf {\bibinfo {volume}
  {26}},\ \bibinfo {pages} {17} (\bibinfo {year} {1981})}\BibitemShut {NoStop}%
\bibitem [{\citenamefont {Dullerud}\ and\ \citenamefont
  {Paganini}(2000)}]{dullerud2000course}%
  \BibitemOpen
  \bibfield  {author} {\bibinfo {author} {\bibfnamefont {G.}~\bibnamefont
  {Dullerud}}\ and\ \bibinfo {author} {\bibfnamefont {F.}~\bibnamefont
  {Paganini}},\ }\href@noop {} {\emph {\bibinfo {title} {Course in Robust
  Control Theory}}}\ (\bibinfo  {publisher} {Springer-Verlag New York},\
  \bibinfo {year} {2000})\BibitemShut {NoStop}%
\bibitem [{\citenamefont {Gugercin}\ and\ \citenamefont
  {Antoulas}(2004)}]{gugercin2004survey}%
  \BibitemOpen
  \bibfield  {author} {\bibinfo {author} {\bibfnamefont {S.}~\bibnamefont
  {Gugercin}}\ and\ \bibinfo {author} {\bibfnamefont {A.~C.}\ \bibnamefont
  {Antoulas}},\ }\href@noop {} {\bibfield  {journal} {\bibinfo  {journal}
  {International Journal of Control}\ }\textbf {\bibinfo {volume} {77}},\
  \bibinfo {pages} {748} (\bibinfo {year} {2004})}\BibitemShut {NoStop}%
\bibitem [{\citenamefont {Zhou}, \citenamefont {D'Souza},\ and\ \citenamefont
  {Cloutier}(1995)}]{zhou1995structurally}%
  \BibitemOpen
  \bibfield  {author} {\bibinfo {author} {\bibfnamefont {K.}~\bibnamefont
  {Zhou}}, \bibinfo {author} {\bibfnamefont {C.}~\bibnamefont {D'Souza}}, \
  and\ \bibinfo {author} {\bibfnamefont {J.~R.}\ \bibnamefont {Cloutier}},\
  }\href@noop {} {\bibfield  {journal} {\bibinfo  {journal} {Systems \& control
  letters}\ }\textbf {\bibinfo {volume} {24}},\ \bibinfo {pages} {235}
  (\bibinfo {year} {1995})}\BibitemShut {NoStop}%
\bibitem [{\citenamefont {Li}\ and\ \citenamefont
  {Paganini}(2005)}]{li2005structured}%
  \BibitemOpen
  \bibfield  {author} {\bibinfo {author} {\bibfnamefont {L.}~\bibnamefont
  {Li}}\ and\ \bibinfo {author} {\bibfnamefont {F.}~\bibnamefont {Paganini}},\
  }\href@noop {} {\bibfield  {journal} {\bibinfo  {journal} {Automatica}\
  }\textbf {\bibinfo {volume} {41}},\ \bibinfo {pages} {145} (\bibinfo {year}
  {2005})}\BibitemShut {NoStop}%
\bibitem [{\citenamefont {Sandberg}\ and\ \citenamefont
  {Murray}(2009)}]{sandberg2009model}%
  \BibitemOpen
  \bibfield  {author} {\bibinfo {author} {\bibfnamefont {H.}~\bibnamefont
  {Sandberg}}\ and\ \bibinfo {author} {\bibfnamefont {R.~M.}\ \bibnamefont
  {Murray}},\ }\href@noop {} {\bibfield  {journal} {\bibinfo  {journal}
  {Optimal Control Applications and Methods}\ }\textbf {\bibinfo {volume}
  {30}},\ \bibinfo {pages} {225} (\bibinfo {year} {2009})}\BibitemShut
  {NoStop}%
\bibitem [{\citenamefont {Scherpen}(1993)}]{scherpen1993balancing}%
  \BibitemOpen
  \bibfield  {author} {\bibinfo {author} {\bibfnamefont {J.~M.}\ \bibnamefont
  {Scherpen}},\ }\href@noop {} {\bibfield  {journal} {\bibinfo  {journal}
  {Systems \& Control Letters}\ }\textbf {\bibinfo {volume} {21}},\ \bibinfo
  {pages} {143} (\bibinfo {year} {1993})}\BibitemShut {NoStop}%
\bibitem [{\citenamefont {Lall}, \citenamefont {Marsden},\ and\ \citenamefont
  {Glava{\v{s}}ki}(2002)}]{lall2002subspace}%
  \BibitemOpen
  \bibfield  {author} {\bibinfo {author} {\bibfnamefont {S.}~\bibnamefont
  {Lall}}, \bibinfo {author} {\bibfnamefont {J.~E.}\ \bibnamefont {Marsden}}, \
  and\ \bibinfo {author} {\bibfnamefont {S.}~\bibnamefont {Glava{\v{s}}ki}},\
  }\href@noop {} {\bibfield  {journal} {\bibinfo  {journal} {International
  journal of robust and nonlinear control}\ }\textbf {\bibinfo {volume} {12}},\
  \bibinfo {pages} {519} (\bibinfo {year} {2002})}\BibitemShut {NoStop}%
\bibitem [{\citenamefont {Krener}(2008)}]{krener2008reduced}%
  \BibitemOpen
  \bibfield  {author} {\bibinfo {author} {\bibfnamefont {A.~J.}\ \bibnamefont
  {Krener}},\ }in\ \href@noop {} {\emph {\bibinfo {booktitle} {Analysis and
  Design of Nonlinear Control Systems}}}\ (\bibinfo  {publisher} {Springer},\
  \bibinfo {year} {2008})\ pp.\ \bibinfo {pages} {41--62}\BibitemShut {NoStop}%
\bibitem [{\citenamefont {Daniels}\ and\ \citenamefont
  {Nemenman}(2014)}]{DanielsN14}%
  \BibitemOpen
  \bibfield  {author} {\bibinfo {author} {\bibfnamefont {B.~C.}\ \bibnamefont
  {Daniels}}\ and\ \bibinfo {author} {\bibfnamefont {I.}~\bibnamefont
  {Nemenman}},\ }\href@noop {} {\bibfield  {journal} {\bibinfo  {journal}
  {arXiv:1404.6283 [q-bio.QM]}\ } (\bibinfo {year} {2014})}\BibitemShut
  {NoStop}%
\bibitem [{\citenamefont {Daniels}\ and\ \citenamefont
  {Nemenman}(2015)}]{DanielsN15}%
  \BibitemOpen
  \bibfield  {author} {\bibinfo {author} {\bibfnamefont {B.~C.}\ \bibnamefont
  {Daniels}}\ and\ \bibinfo {author} {\bibfnamefont {I.}~\bibnamefont
  {Nemenman}},\ }\href@noop {} {\bibfield  {journal} {\bibinfo  {journal} {PLOS
  ONE}\ } (\bibinfo {year} {in press, 2015})}\BibitemShut {NoStop}%
\bibitem [{\citenamefont {Transtrum}\ and\ \citenamefont
  {Qiu}(2014)}]{transtrum2014model}%
  \BibitemOpen
  \bibfield  {author} {\bibinfo {author} {\bibfnamefont {M.~K.}\ \bibnamefont
  {Transtrum}}\ and\ \bibinfo {author} {\bibfnamefont {P.}~\bibnamefont
  {Qiu}},\ }\href@noop {} {\bibfield  {journal} {\bibinfo  {journal} {Physical
  Review Letters}\ }\textbf {\bibinfo {volume} {113}},\ \bibinfo {pages}
  {098701} (\bibinfo {year} {2014})}\BibitemShut {NoStop}%
\bibitem [{\citenamefont {Transtrum}, \citenamefont {Hart},\ and\ \citenamefont
  {Qiu}(2014)}]{transtrum2014information}%
  \BibitemOpen
  \bibfield  {author} {\bibinfo {author} {\bibfnamefont {M.~K.}\ \bibnamefont
  {Transtrum}}, \bibinfo {author} {\bibfnamefont {G.}~\bibnamefont {Hart}}, \
  and\ \bibinfo {author} {\bibfnamefont {P.}~\bibnamefont {Qiu}},\ }\href@noop
  {} {\bibfield  {journal} {\bibinfo  {journal} {arXiv preprint
  arXiv:1409.6203}\ } (\bibinfo {year} {2014})}\BibitemShut {NoStop}%
\bibitem [{\citenamefont {Apgar}\ \emph {et~al.}(2010)\citenamefont {Apgar},
  \citenamefont {Witmer}, \citenamefont {White},\ and\ \citenamefont
  {Tidor}}]{ApgarWWT10}%
  \BibitemOpen
  \bibfield  {author} {\bibinfo {author} {\bibfnamefont {J.~F.}\ \bibnamefont
  {Apgar}}, \bibinfo {author} {\bibfnamefont {D.~K.}\ \bibnamefont {Witmer}},
  \bibinfo {author} {\bibfnamefont {F.~M.}\ \bibnamefont {White}}, \ and\
  \bibinfo {author} {\bibfnamefont {B.}~\bibnamefont {Tidor}},\ }\href@noop {}
  {\bibfield  {journal} {\bibinfo  {journal} {Mol. Biosyst.}\ }\textbf
  {\bibinfo {volume} {6}},\ \bibinfo {pages} {1890} (\bibinfo {year}
  {2010})}\BibitemShut {NoStop}%
\bibitem [{\citenamefont {Vilela}\ \emph {et~al.}(2009)\citenamefont {Vilela},
  \citenamefont {Vinga}, \citenamefont {Maia}, \citenamefont {Voit},\ and\
  \citenamefont {Almeida}}]{VilelaVMVA09}%
  \BibitemOpen
  \bibfield  {author} {\bibinfo {author} {\bibfnamefont {M.}~\bibnamefont
  {Vilela}}, \bibinfo {author} {\bibfnamefont {S.}~\bibnamefont {Vinga}},
  \bibinfo {author} {\bibfnamefont {M.~A.}\ \bibnamefont {Maia}}, \bibinfo
  {author} {\bibfnamefont {E.~O.}\ \bibnamefont {Voit}}, \ and\ \bibinfo
  {author} {\bibfnamefont {J.~S.}\ \bibnamefont {Almeida}},\ }\href@noop {}
  {\bibfield  {journal} {\bibinfo  {journal} {BMC systems biology}\ }\textbf
  {\bibinfo {volume} {3}},\ \bibinfo {pages} {47} (\bibinfo {year}
  {2009})}\BibitemShut {NoStop}%
\bibitem [{\citenamefont {Erguler}\ and\ \citenamefont
  {Stumpf}(2011)}]{ErgulerS11}%
  \BibitemOpen
  \bibfield  {author} {\bibinfo {author} {\bibfnamefont {K.}~\bibnamefont
  {Erguler}}\ and\ \bibinfo {author} {\bibfnamefont {M.~P.~H.}\ \bibnamefont
  {Stumpf}},\ }\href@noop {} {\bibfield  {journal} {\bibinfo  {journal}
  {Molecular BioSystems}\ }\textbf {\bibinfo {volume} {7}},\ \bibinfo {pages}
  {1593} (\bibinfo {year} {2011})}\BibitemShut {NoStop}%
\bibitem [{\citenamefont {Transtrum}\ and\ \citenamefont
  {Qiu}(2012)}]{TranstrumQ12}%
  \BibitemOpen
  \bibfield  {author} {\bibinfo {author} {\bibfnamefont {M.}~\bibnamefont
  {Transtrum}}\ and\ \bibinfo {author} {\bibfnamefont {P.}~\bibnamefont
  {Qiu}},\ }\href {\doibase 10.1186/1471-2105-13-181} {\bibfield  {journal}
  {\bibinfo  {journal} {BMC Bioinformatics}\ }\textbf {\bibinfo {volume}
  {13}},\ \bibinfo {pages} {181} (\bibinfo {year} {2012})}\BibitemShut
  {NoStop}%
\bibitem [{\citenamefont {Chachra}, \citenamefont {Transtrum},\ and\
  \citenamefont {Sethna}(2011)}]{ChachraTS11}%
  \BibitemOpen
  \bibfield  {author} {\bibinfo {author} {\bibfnamefont {R.}~\bibnamefont
  {Chachra}}, \bibinfo {author} {\bibfnamefont {M.~K.}\ \bibnamefont
  {Transtrum}}, \ and\ \bibinfo {author} {\bibfnamefont {J.~P.}\ \bibnamefont
  {Sethna}},\ }\href {\doibase 10.1039/C1MB05046J} {\bibfield  {journal}
  {\bibinfo  {journal} {Mol. BioSyst.}\ ,\ } (\bibinfo {year}
  {2011})}\BibitemShut {NoStop}%
\bibitem [{\citenamefont {Lee}\ \emph {et~al.}(2008)\citenamefont {Lee},
  \citenamefont {Salic}, \citenamefont {Kr{\:u}ger}, \citenamefont {Heinrich},\
  and\ \citenamefont {Kirschner}}]{LeeSKHK03}%
  \BibitemOpen
  \bibfield  {author} {\bibinfo {author} {\bibfnamefont {E.}~\bibnamefont
  {Lee}}, \bibinfo {author} {\bibfnamefont {A.}~\bibnamefont {Salic}}, \bibinfo
  {author} {\bibfnamefont {R.}~\bibnamefont {Kr{\:u}ger}}, \bibinfo {author}
  {\bibfnamefont {R.}~\bibnamefont {Heinrich}}, \ and\ \bibinfo {author}
  {\bibfnamefont {M.~W.}\ \bibnamefont {Kirschner}},\ }\href@noop {} {\bibfield
   {journal} {\bibinfo  {journal} {PLoS Biology}\ }\textbf {\bibinfo {volume}
  {1}},\ \bibinfo {pages} {e10} (\bibinfo {year} {2008})}\BibitemShut {NoStop}%
\bibitem [{\citenamefont {Ringner}(2008)}]{Ringner08}%
  \BibitemOpen
  \bibfield  {author} {\bibinfo {author} {\bibfnamefont {M.}~\bibnamefont
  {Ringner}},\ }\href {\doibase 10.1038/nbt0308-303} {\bibfield  {journal}
  {\bibinfo  {journal} {Nat Biotech}\ }\textbf {\bibinfo {volume} {26}},\
  \bibinfo {pages} {303} (\bibinfo {year} {2008})}\BibitemShut {NoStop}%
\bibitem [{\citenamefont {Levenberg}(1944)}]{Levenberg1944}%
  \BibitemOpen
  \bibfield  {author} {\bibinfo {author} {\bibfnamefont {K.}~\bibnamefont
  {Levenberg}},\ }\href@noop {} {\bibfield  {journal} {\bibinfo  {journal}
  {Quart. Appl. Math}\ }\textbf {\bibinfo {volume} {2}},\ \bibinfo {pages}
  {164} (\bibinfo {year} {1944})}\BibitemShut {NoStop}%
\bibitem [{\citenamefont {Marquardt}(1963)}]{Marquardt1963}%
  \BibitemOpen
  \bibfield  {author} {\bibinfo {author} {\bibfnamefont {D.}~\bibnamefont
  {Marquardt}},\ }\href@noop {} {\bibfield  {journal} {\bibinfo  {journal}
  {Journal of the Society for Industrial and Applied Mathematics}\ }\textbf
  {\bibinfo {volume} {11}},\ \bibinfo {pages} {431} (\bibinfo {year}
  {1963})}\BibitemShut {NoStop}%
\bibitem [{\citenamefont {Press}\ \emph {et~al.}(2007)\citenamefont {Press},
  \citenamefont {Teukolsky}, \citenamefont {Vetterling},\ and\ \citenamefont
  {Flannery}}]{Press2007}%
  \BibitemOpen
  \bibfield  {author} {\bibinfo {author} {\bibfnamefont {W.}~\bibnamefont
  {Press}}, \bibinfo {author} {\bibfnamefont {S.~A.}\ \bibnamefont
  {Teukolsky}}, \bibinfo {author} {\bibfnamefont {W.~T.}\ \bibnamefont
  {Vetterling}}, \ and\ \bibinfo {author} {\bibfnamefont {B.~P.}\ \bibnamefont
  {Flannery}},\ }\href@noop {} {\emph {\bibinfo {title} {Numerical recipes: the
  art of scientific computing,}}}\ (\bibinfo  {publisher} {Cambridge University
  Press},\ \bibinfo {year} {2007})\BibitemShut {NoStop}%
\bibitem [{\citenamefont {Transtrum}\ and\ \citenamefont
  {Sethna}()}]{TranstrumSnnb}%
  \BibitemOpen
  \bibfield  {author} {\bibinfo {author} {\bibfnamefont {M.~K.}\ \bibnamefont
  {Transtrum}}\ and\ \bibinfo {author} {\bibfnamefont {J.~P.}\ \bibnamefont
  {Sethna}},\ }\href@noop {} {\bibinfo  {journal} {(manuscript in revision),
  \url{http://arxiv.org/abs/1201.5885}}\ }\BibitemShut {NoStop}%
\bibitem [{\citenamefont {Transtrum}\ and\ \citenamefont
  {Sethna}(2011)}]{url:GeodesicAccelerationSourceForge}%
  \BibitemOpen
\bibfield  {journal} {  }\bibfield  {author} {\bibinfo {author} {\bibfnamefont
  {M.}~\bibnamefont {Transtrum}}\ and\ \bibinfo {author} {\bibfnamefont
  {J.~P.}\ \bibnamefont {Sethna}},\ }\href@noop {} {\enquote {\bibinfo {title}
  {geodesiclm},}\ }\bibinfo {howpublished}
  {\url{http://sourceforge.net/projects/geodesiclm/}} (\bibinfo {year}
  {2011})\BibitemShut {NoStop}%
\bibitem [{\citenamefont {Daniels}\ \emph {et~al.}(2008)\citenamefont
  {Daniels}, \citenamefont {Chen}, \citenamefont {Sethna}, \citenamefont
  {Gutenkunst},\ and\ \citenamefont {Myers}}]{DanielsCSGM08}%
  \BibitemOpen
  \bibfield  {author} {\bibinfo {author} {\bibfnamefont {B.~C.}\ \bibnamefont
  {Daniels}}, \bibinfo {author} {\bibfnamefont {Y.~J.}\ \bibnamefont {Chen}},
  \bibinfo {author} {\bibfnamefont {J.~P.}\ \bibnamefont {Sethna}}, \bibinfo
  {author} {\bibfnamefont {R.~N.}\ \bibnamefont {Gutenkunst}}, \ and\ \bibinfo
  {author} {\bibfnamefont {C.~R.}\ \bibnamefont {Myers}},\ }\href@noop {}
  {\bibfield  {journal} {\bibinfo  {journal} {Current Opinion In
  Biotechnology}\ }\textbf {\bibinfo {volume} {19}},\ \bibinfo {pages} {389}
  (\bibinfo {year} {2008})}\BibitemShut {NoStop}%
\bibitem [{\citenamefont {Wagner}(2005)}]{Wagner05}%
  \BibitemOpen
  \bibfield  {author} {\bibinfo {author} {\bibfnamefont {A.}~\bibnamefont
  {Wagner}},\ }\href@noop {} {\emph {\bibinfo {title} {Robustness and
  Evolvability in Living Systems}}}\ (\bibinfo  {publisher} {Princeton
  University Press},\ \bibinfo {year} {2005})\BibitemShut {NoStop}%
\bibitem [{\citenamefont {Cutler}\ and\ \citenamefont
  {Breiman}(1994)}]{Cutler94}%
  \BibitemOpen
  \bibfield  {author} {\bibinfo {author} {\bibfnamefont {A.}~\bibnamefont
  {Cutler}}\ and\ \bibinfo {author} {\bibfnamefont {L.}~\bibnamefont
  {Breiman}},\ }\href@noop {} {\bibfield  {journal} {\bibinfo  {journal}
  {Technometrics}\ }\textbf {\bibinfo {volume} {36}},\ \bibinfo {pages} {338}
  (\bibinfo {year} {1994})}\BibitemShut {NoStop}%
\bibitem [{\citenamefont {Mørup}\ and\ \citenamefont
  {Hansen}(2012)}]{MorupH12}%
  \BibitemOpen
  \bibfield  {author} {\bibinfo {author} {\bibfnamefont {M.}~\bibnamefont
  {Mørup}}\ and\ \bibinfo {author} {\bibfnamefont {L.~K.}\ \bibnamefont
  {Hansen}},\ }\href {\doibase http://dx.doi.org/10.1016/j.neucom.2011.06.033}
  {\bibfield  {journal} {\bibinfo  {journal} {Neurocomputing}\ }\textbf
  {\bibinfo {volume} {80}},\ \bibinfo {pages} {54 } (\bibinfo {year} {2012})},\
  \bibinfo {note} {special Issue on Machine Learning for Signal Processing
  2010}\BibitemShut {NoStop}%
\bibitem [{\citenamefont {Thurau}, \citenamefont {Kersting},\ and\
  \citenamefont {Bauckhage}(2009)}]{ThurauKB09}%
  \BibitemOpen
  \bibfield  {author} {\bibinfo {author} {\bibfnamefont {C.}~\bibnamefont
  {Thurau}}, \bibinfo {author} {\bibfnamefont {K.}~\bibnamefont {Kersting}}, \
  and\ \bibinfo {author} {\bibfnamefont {C.}~\bibnamefont {Bauckhage}},\ }in\
  \href {\doibase 10.1109/ICDM.2009.55} {\emph {\bibinfo {booktitle} {Data
  Mining, 2009. ICDM '09. Ninth IEEE International Conference on}}}\ (\bibinfo
  {year} {2009})\ pp.\ \bibinfo {pages} {523--532}\BibitemShut {NoStop}%
\bibitem [{\citenamefont {Koren}, \citenamefont {Bell},\ and\ \citenamefont
  {Volinsky}(2009)}]{KorenBV09}%
  \BibitemOpen
  \bibfield  {author} {\bibinfo {author} {\bibfnamefont {Y.}~\bibnamefont
  {Koren}}, \bibinfo {author} {\bibfnamefont {R.}~\bibnamefont {Bell}}, \ and\
  \bibinfo {author} {\bibfnamefont {C.}~\bibnamefont {Volinsky}},\ }\href@noop
  {} {\bibfield  {journal} {\bibinfo  {journal} {Computer}\ }\textbf {\bibinfo
  {volume} {42}},\ \bibinfo {pages} {30} (\bibinfo {year} {2009})}\BibitemShut
  {NoStop}%
\bibitem [{\citenamefont {Vincent}\ \emph {et~al.}(2008)\citenamefont
  {Vincent}, \citenamefont {Larochelle}, \citenamefont {Bengio},\ and\
  \citenamefont {Manzagol}}]{VincentLBM08}%
  \BibitemOpen
  \bibfield  {author} {\bibinfo {author} {\bibfnamefont {P.}~\bibnamefont
  {Vincent}}, \bibinfo {author} {\bibfnamefont {H.}~\bibnamefont {Larochelle}},
  \bibinfo {author} {\bibfnamefont {Y.}~\bibnamefont {Bengio}}, \ and\ \bibinfo
  {author} {\bibfnamefont {P.-A.}\ \bibnamefont {Manzagol}},\ }in\ \href
  {\doibase 10.1145/1390156.1390294} {\emph {\bibinfo {booktitle} {Proceedings
  of the 25th International Conference on Machine Learning}}},\ \bibinfo
  {series and number} {ICML '08}\ (\bibinfo  {publisher} {ACM},\ \bibinfo
  {address} {New York, NY, USA},\ \bibinfo {year} {2008})\ pp.\ \bibinfo
  {pages} {1096--1103}\BibitemShut {NoStop}%
\bibitem [{\citenamefont {Hayden}, \citenamefont {Alemi},\ and\ \citenamefont
  {Sethn}()}]{HaydenASnn}%
  \BibitemOpen
  \bibfield  {author} {\bibinfo {author} {\bibfnamefont {L.~X.}\ \bibnamefont
  {Hayden}}, \bibinfo {author} {\bibfnamefont {A.~A.}\ \bibnamefont {Alemi}}, \
  and\ \bibinfo {author} {\bibfnamefont {J.~P.}\ \bibnamefont {Sethn}},\
  }\href@noop {} {\bibinfo  {journal} {(work in progress)}\ }\BibitemShut
  {NoStop}%
\bibitem [{\citenamefont {Sethna}(2006)}]{Sethna06}%
  \BibitemOpen
\bibfield  {journal} {  }\bibfield  {author} {\bibinfo {author} {\bibfnamefont
  {J.~P.}\ \bibnamefont {Sethna}},\ }\href@noop {} {\emph {\bibinfo {title}
  {Statistical Mechanics: Entropy, Order Parameters, and Complexity,
  \url{http://www.physics.cornell.edu/sethna/StatMech/}}}}\ (\bibinfo
  {publisher} {Oxford University Press},\ \bibinfo {address} {Oxford},\
  \bibinfo {year} {2006})\BibitemShut {NoStop}%
\bibitem [{\citenamefont {Casey}\ \emph {et~al.}(2007)\citenamefont {Casey},
  \citenamefont {Baird}, \citenamefont {Feng}, \citenamefont {Gutenkunst},
  \citenamefont {Waterfall}, \citenamefont {Myers}, \citenamefont {Brown},
  \citenamefont {Cerione},\ and\ \citenamefont {Sethna}}]{CaseyBFGWMBCS07}%
  \BibitemOpen
  \bibfield  {author} {\bibinfo {author} {\bibfnamefont {F.~P.}\ \bibnamefont
  {Casey}}, \bibinfo {author} {\bibfnamefont {D.}~\bibnamefont {Baird}},
  \bibinfo {author} {\bibfnamefont {Q.}~\bibnamefont {Feng}}, \bibinfo {author}
  {\bibfnamefont {R.~N.}\ \bibnamefont {Gutenkunst}}, \bibinfo {author}
  {\bibfnamefont {J.~J.}\ \bibnamefont {Waterfall}}, \bibinfo {author}
  {\bibfnamefont {C.~R.}\ \bibnamefont {Myers}}, \bibinfo {author}
  {\bibfnamefont {K.~S.}\ \bibnamefont {Brown}}, \bibinfo {author}
  {\bibfnamefont {R.~A.}\ \bibnamefont {Cerione}}, \ and\ \bibinfo {author}
  {\bibfnamefont {J.~P.}\ \bibnamefont {Sethna}},\ }\href@noop {} {\bibfield
  {journal} {\bibinfo  {journal} {Iet Systems Biology}\ }\textbf {\bibinfo
  {volume} {1}},\ \bibinfo {pages} {190} (\bibinfo {year} {2007})}\BibitemShut
  {NoStop}%
\bibitem [{\citenamefont {Kirschner}\ and\ \citenamefont
  {Gerhart}(1998)}]{KirschnerG98}%
  \BibitemOpen
  \bibfield  {author} {\bibinfo {author} {\bibfnamefont {M.}~\bibnamefont
  {Kirschner}}\ and\ \bibinfo {author} {\bibfnamefont {J.}~\bibnamefont
  {Gerhart}},\ }\href {http://www.pnas.org/cgi/content/abstract/95/15/8420}
  {\bibfield  {journal} {\bibinfo  {journal} {Proceedings of the National
  Academy of Sciences}\ }\textbf {\bibinfo {volume} {95}},\ \bibinfo {pages}
  {8420} (\bibinfo {year} {1998})}\BibitemShut {NoStop}%
\bibitem [{\citenamefont {Hartwell}\ \emph {et~al.}(1999)\citenamefont
  {Hartwell}, \citenamefont {Hopfield}, \citenamefont {Leibler},\ and\
  \citenamefont {Murray}}]{HartwellHLM99}%
  \BibitemOpen
  \bibfield  {author} {\bibinfo {author} {\bibfnamefont {L.~H.}\ \bibnamefont
  {Hartwell}}, \bibinfo {author} {\bibfnamefont {J.~J.}\ \bibnamefont
  {Hopfield}}, \bibinfo {author} {\bibfnamefont {S.}~\bibnamefont {Leibler}}, \
  and\ \bibinfo {author} {\bibfnamefont {A.~W.}\ \bibnamefont {Murray}},\
  }\href@noop {} {\bibfield  {journal} {\bibinfo  {journal} {Nature}\ }\textbf
  {\bibinfo {volume} {402}},\ \bibinfo {pages} {C47} (\bibinfo {year}
  {1999})}\BibitemShut {NoStop}%
\bibitem [{\citenamefont {Kashtan}\ and\ \citenamefont
  {Alon}(2005)}]{KashtanA05}%
  \BibitemOpen
  \bibfield  {author} {\bibinfo {author} {\bibfnamefont {N.}~\bibnamefont
  {Kashtan}}\ and\ \bibinfo {author} {\bibfnamefont {U.}~\bibnamefont {Alon}},\
  }\href {http://www.pnas.org/cgi/content/abstract/102/39/13773} {\bibfield
  {journal} {\bibinfo  {journal} {Proceedings of the National Academy of
  Sciences}\ }\textbf {\bibinfo {volume} {102}},\ \bibinfo {pages} {13773}
  (\bibinfo {year} {2005})}\BibitemShut {NoStop}%
\bibitem [{\citenamefont {Clune}, \citenamefont {Mouret},\ and\ \citenamefont
  {Lipson}(2013)}]{CluneML13}%
  \BibitemOpen
  \bibfield  {author} {\bibinfo {author} {\bibfnamefont {J.}~\bibnamefont
  {Clune}}, \bibinfo {author} {\bibfnamefont {J.-B.}\ \bibnamefont {Mouret}}, \
  and\ \bibinfo {author} {\bibfnamefont {H.}~\bibnamefont {Lipson}},\ }\href
  {\doibase 10.1098/rspb.2012.2863} {\bibfield  {journal} {\bibinfo  {journal}
  {Proceedings of the Royal Society of London B: Biological Sciences}\ }\textbf
  {\bibinfo {volume} {280}} (\bibinfo {year} {2013}),\
  10.1098/rspb.2012.2863}\BibitemShut {NoStop}%
\end{thebibliography}%
